\documentclass[12pt,a4paper]{article}
\usepackage{amsmath,amsfonts,amssymb,amscd,amsthm}
\usepackage{mathdots}
\usepackage{color}
\usepackage[bottom=0.5in,top=0.5in,left=0.5in,right=0.5in]{geometry}
\usepackage{hyperref}

\numberwithin{equation}{section}

\newcommand{\p}{\partial}
\DeclareMathOperator{\rk}{rk}

\newtheorem{theorem}{Theorem}[section]
\newtheorem{lemma}[theorem]{Lemma}
\newtheorem{corollary}[theorem]{Corollary}
\newtheorem{proposition}[theorem]{Proposition}
\newtheorem{conjecture}{Conjecture}

\newtheorem*{algorithm*}{Algorithm}

\theoremstyle{definition}
\newtheorem{definition}[theorem]{Definition}

\theoremstyle{remark}
\newtheorem{remark}[theorem]{Remark}

\numberwithin{equation}{section}

\newcommand{\todo}[1][\null]{\ensuremath{\clubsuit}}

\newcommand{\noprint}[1]{}
\newcommand{\checked}[1][\null]{\ensuremath{\boldsymbol{\surd}}}

\newcommand*{\pd}
[2]{\mathchoice{\frac{\partial#1}{\partial#2}}
  {\partial#1/\partial#2}{\partial#1/\partial#2}
  {\partial#1/\partial#2}}

\newcommand*{\fd}
[2]{\mathchoice{\frac{\delta#1}{\delta#2}}
  {\delta #1/\delta#2}{\delta#1/\delta#2}{\delta#1/\delta#2}}

\newcommand{\rn}{\mathbb{R}^N}

\DeclareMathOperator{\Span}{Span}
\allowdisplaybreaks[3]

\begin{document}

\title{On the geometry of WDVV equations
  \\
  and their Hamiltonian formalism
  \\
  in arbitrary dimension}
\author{S. Opanasenko\thanks{This
  research has been partially supported by the research project Mathematical
  Methods in Non Linear Physics (MMNLP) by the Commissione Scientifica
  Nazionale -- Gruppo 4 -- Fisica Teorica of the Istituto Nazionale di Fisica
  Nucleare (INFN), the Department of Mathematics and Physics ``E. De Giorgi''
  of the Universit\`a del Salento, GNFM of the Istituto Nazionale di Alta
  Matematica (INdAM), PRIN 2022TEB52W \emph{The charm of integrability: from
    nonlinear waves to random matrices}, ICSC -- Centro Nazionale di Ricerca in
  High Performance Computing, Big Data and Quantum Computing, funded by
  European Union -- NextGenerationEU and PRIN2020 F3NCPX``Mathematics for
  Industry 4.0''.}
  \\
\small  INFN, sezione di Lecce
  \\
  \small  via per Arnesano, 73100 Lecce, Italy
  \\
  \small  and Institute of Mathematics of NAS of Ukraine
  \\
  \small 3 Te\-re\-shchen\-kiv\-ska Str., 01024 Kyiv, Ukraine
  \\
  \small \texttt{stanislav.opanasenko@le.infn.it}
  \and R. Vitolo$^*$
  \\
  \small Dipartimento di Matematica e Fisica ``E. De Giorgi'' Universit\`a del
  Salento
  \\
  \small via per Arnesano, 73100 Lecce, Italy and INFN sezione di Lecce
  \\
  \small \texttt{raffaele.vitolo@unisalento.it}
}

\maketitle

\begin{abstract}
  It is known that in low dimensions WDVV equations can be rewritten as
  commuting quasilinear bi-Hamiltonian systems.  We extend some of these
  results to arbitrary dimension $N$ and arbitrary scalar product $\eta$.  In
  particular, we show that suitable subsets of WDVV equations can be
  interpreted as a set of linear line congruences in suitable Pl\"ucker
  embeddings.  This form leads to their representation as $N-2$ Hamiltonian
  systems of conservation laws.  Moreover, we show that WDVV equations can be
  reduced to an orthonomic form, which is also passive in low dimensions
  $N\leq 5$.  This also leads to the commutativity of the Hamiltonian systems
  of conservation laws ($N\leq 5$), after which we can find a solution of the
  WDVV equations from a joint solution of the Hamiltonian systems. Finally, we
  conjecture that passivity holds in all dimensions.
\end{abstract}

\tableofcontents

\section{Introduction}
\label{page:introduction}

\subsection{WDVV equations}

Witten--Dijkgraaf--Verlinde--Verlinde (WDVV) equations arose within the
framework of topological field theory~\cite{DIJKGRAAF199159,WITTEN1990281}.
Later, Dubrovin~\cite{D96} found striking connections between them and the
theory of Integrable Systems, which led him to introduce Frobenius manifolds.
The \emph{potential} of a Frobenius manifold is a solution of the WDVV
equations.  Nowadays, Frobenius Manifolds are a very active research subject
among mathematicians: see, \emph{e.g.}, the recent works
\cite{Alkadhem_2021,arsie23:_semis_flat_f,Basalaev_2021,
  brini2025dubrovindualitymirrorsymmetry, Cao2021,cotti2020degenerate,
  Ferapontov2021,iglesias22:_bi_hamil_recur,liu25:_gener_froben,
  StedmanStrachanJMP2021,vollmer:_manif_hamil,zinger2020real}.  Another
remarkable achievement was the discovery in~\cite{kontsevich94:_gromov} that a
recursive formula for Gromov--Witten invariants for~$\mathbb{P}^2$ could be
found by solving WDVV equations.  WDVV equations are also actively studied
within the community of Theoretical Physicists, with ongoing research on
supersymmetric quantum
mechanics~\cite{Antoniou2019,galajinsky24:_remar,Kozyrev2018}, topological
quantum field theory~\cite{Gomez2021}, string
theory~\cite{Belavin2014,semenyakin22:_topol} and supersymmetric gauge
theory~\cite{jockers22_bps}.

We introduce WDVV equations following the presentation
in~\cite{dubrovin06:_encyc_mathem_physic}
by defining the potential $F:\mathbb{R}^N\to\mathbb R$ such that
\begin{enumerate}
\item $\eta_{\alpha\beta}:=F_{1\alpha\beta}$ is a constant symmetric
  nondegenerate matrix, which we interpret as a metric $\eta$
  on~$\mathbb{R}^N$;
\item $c^\gamma_{\alpha\beta} := \eta^{\gamma\epsilon} F_{\epsilon\alpha\beta}$
  (here $(\eta^{\alpha\beta})=(\eta_{\alpha\beta})^{-1}$)
  are the structure constants of an associative algebra;\label{item:2}
\item $F$ is quasihomogeneous:
  $F(c^{d_1}t^1,\ldots,c^{d_N}t^N) = c^{d_F}F(t^1,\ldots,t^N)$.\label{item:3}
\end{enumerate}
Here and below,
\begin{displaymath}
  F_{\alpha\beta\gamma}:=\frac{\p^3F}{\p t^\alpha \p t^\beta \p t^\gamma}.
\end{displaymath}
If $e_1$,\dots, $e_N$ is the basis of $\mathbb{R}^N$ then the algebra operation
is $e_\alpha\cdot e_\beta = c^\gamma_{\alpha\beta}(\mathbf{t})e_\gamma$ with
$e_1$ being the unity.
The associativity condition for this operation takes the form of the
following system of PDEs
\begin{equation}\label{eq:5}
  S_{\alpha\beta\gamma\nu}:=
  \eta ^{\lambda\mu}(F_{\lambda\alpha\beta }F_{\mu\gamma\nu}
  - F_{\lambda\alpha\nu}F_{\mu\gamma\beta})=0.
\end{equation}
It can be reduced to a system with the unknown function
$f=f(t^2,\ldots, t^N)$, as the above requirements completely specify the
functional dependence of $F$ on $t^1$ \emph{up to second degree polynomials},
\begin{equation*}
  F(t^1,\dots,t^N)=\frac{1}{6}\eta_{11}(t^1)^3 +
  \frac{1}{2}\sum_{k>1}\eta_{1k}t^k(t^1)^2 + \frac{1}{2}\sum_{k,s>1}
  \eta_{sk}t^st^kt^1 + f(t^2,\ldots,t^N).
\end{equation*}
The WDVV equations are the overdetermined nonlinear system of PDEs \eqref{eq:5}
in one unknown function~$f$ of $N-1$~independent variables, which specify the
associativity of the above operation, together with the quasihomogeneity
equation in~\eqref{item:3}. We will refer to the system~\eqref{eq:5} as the
WDVV equations, with or without the additional quasihomogeneity equation.

Two choices determine WDVV equations: the dimension~$N$ and the metric $\eta$.
If $N=1$ the WDVV equation is trivial, and if $N=2$ the associativity condition
is trivial, leaving only the quasihomogeneity requirement (see \cite{D96} for
more details). So, the interesting cases for us are $N\geq 3$ and arbitrary
$\eta$.

It is proved in~\cite{D96} that, in the case of distinct quasihomogeneity
weights, unity-preserving linear transformations of $\mathbb{R}^N$ bring the
metric~$\eta$ to the canonical forms $\eta^{(1)}=(\eta^{(1)}_{\alpha\beta})$
when $\eta_{11}=0$, with
\begin{equation*}
  \eta^{(1)}=
  \begin{pmatrix}
    0 & & 1
    \\
    & \iddots
    &
    \\
    1 & & 0
  \end{pmatrix}
\end{equation*}
and $\eta^{(2)}=(\eta^{(2)}_{\alpha\beta})$ when $\eta_{11}\neq 0$, with
\begin{equation*}
  \eta^{(2)}=
  \begin{pmatrix}
    \mu & & 1
    \\
    & \iddots
    &
    \\
    1 & & 0
  \end{pmatrix}
\end{equation*}
where $\mu\neq 0$.

\subsection{First problem: determining nontrivial
  WDVV equations}
\label{sec:first-probl-determ}

The first nontrivial WDVV equation appears in dimension $N=3$ \cite{D96}:
it is a single equation on $f=f(x,t)$, where $x=t^2$, $t=t^3$.
In particular, for the canonical form $\eta=\eta^{(1)}$ it is
\begin{equation}\label{eq:WDVV:simple}
  f_{ttt}=f_{xxt}^2 - f_{xxx}f_{xtt}.
\end{equation}

In dimension $N=4$, for the canonical form $\eta=\eta^{(1)}$, after relabeling
the independent variables as $x:=t^2$, $y:=t^3$, $z:=t^4$, we have the WDVV
system of six equations~\cite{D96}
\begin{equation}\label{eq:34}
  \begin{split}
    & -2f_{xyz}-f_{xyy}f_{xxy}+f_{yyy}f_{xxx} = 0,
    \\
    & -f_{xzz}-f_{xyy}f_{xxz}+f_{yyz}f_{xxx} = 0,
    \\
    & -2f_{xyz}f_{xxz}+f_{xzz}f_{xxy}+f_{yzz}f_{xxx} = 0,
    \\
    & f_{xxy}f_{yyz} - f_{xxz}f_{yyy} + f_{yzz} = 0,
    \\
    & f_{zzz}-(f_{xyz})^2+f_{xzz}f_{xyy}-f_{yyz}f_{xxz}+f_{yzz}f_{xxy} = 0,
    \\
    & f_{yyy}f_{xzz}-2f_{yyz}f_{xyz}+f_{yzz}f_{xyy} = 0.
  \end{split}
\end{equation}
We observe a feature of the above system that is implicitly present
in~\cite{ferapontov96:_hamil}. If we choose one independent variable, say~$x$
(without loss of generality), and introduce new field variables
\begin{gather}\label{eq:15}
  u^{1}=f_{xxx},\ u^{2}=f_{xxy},\ u^{3}=f_{xyy},\ u^{4}=f_{xxz},\
  u^{5}=f_{xyz},\ u^{6}=f_{xzz},
\end{gather}
then we can solve the WDVV system~\eqref{eq:34} with respect to the remaining
variables $f_{yyy}$, $f_{yyz}$, $f_{yzz}$, $f_{zzz}$ ($x$-free derivatives) in
an explicit way.  More precisely, in the system~\eqref{eq:34} all equations are
linear with respect to the $x$-free derivatives.  After solving a subsystem of
any four equations, there are two remaining equations, both of which are
identically satisfied by means of the values of the $x$-free derivatives
obtained in terms of the $x$-derivative variables~\eqref{eq:15}.  So, the WDVV
system~\eqref{eq:34} is actually equivalent to the system
\begin{equation*}
  \begin{split}
    &f_{yyy}=\frac{2u^{5}+u^{2}u^{3}}{u^{1}},
    \quad
    f_{yyz}=\frac{u^{3}u^{4}+u^{6}}{u^{1}},
    \quad
    f_{yzz}=\frac{2u^{4}u^{5}-u^{2}u^{6}}{u^{1}},
    \\
    &f_{zzz}=(u^{5})^{2}-u^{3}u^{6}+ \frac{
      u^{3}(u^{4})^{2}+u^{4}u^{6}-2u^{2}u^{4}u^{5}+(u^{2})^{2}u^{6}}{u^{1}}.
  \end{split}
\end{equation*}
Of course, the above solution is valid wherever the denominator does not
vanish. That is an open dense subset of the space of dependent variables.

Note that a similar result can be obtained if we choose $y$ or $z$.

\textbf{Problem:} can we generalize the above feature to WDVV equations for
generic metrics $\eta$ and dimensions $N$?  More precisely, in the general
case, having chosen an arbitrary independent variable $t^p$, is it possible to
reduce the WDVV system to a system that consists of equations for $t^p$-free
derivative variables in terms of $t^p$-derivative variables?
After having achieved the above result, what about compatibility of the
``reduced'' WDVV system?

\begin{remark}
  The above problem, \emph{i.e.} reducibility and compatibility of WDVV
  equations, makes sense only when there is more than one WDVV equation. That's
  why the first generic case, and the example that we will work out in full
  details, is in dimension $N=4$.
\end{remark}

\subsection{Second problem: finding a Hamiltonian formulation}
\label{sec:second-probl-find}

The integrability of the WDVV equations was considered in \cite{D96}, where a
Lax pair for the WDVV equations was provided. The geometric nature of the WDVV
equations makes the Lax pair nature quite transparent: since the WDVV equations
are conditions of flatness of a Levi-Civita connection (see also
\cite{dubrovin98:_flat_froben}), the (linear) equation of flat
vectors/covectors can be interpreted as its Lax pair.

Moreover, it was shown in~\cite{FGMN97} that the simplest WDVV
equation~\eqref{eq:WDVV:simple} ($N=3$, $\eta=\eta^{(1)}$) is integrable also
in the sense that it is bi-Hamiltonian.  Since the Hamiltonian formalism for
PDEs is developed mostly for evolutionary equations, the Authors used the idea
of Mokhov~\cite{mokhov95:_sympl_poiss} to rewrite the
equation~\eqref{eq:WDVV:simple} as a first-order quasilinear system (or a
hydrodynamic-type system) of PDEs.  More precisely, introducing the coordinates
$u^1=f_{xxx}$, $u^2=f_{xxt}$, $u^3=f_{xtt}$, $u^4=f_{ttt}$ we observe that
compatibility conditions for such coordinates are
\begin{equation*}
  u^1_t = u^2_x,\quad u^2_t = u^3_x,\quad u^3_t = u^4_x.
\end{equation*}
With the help of the equation~\eqref{eq:WDVV:simple}, the above compatibility conditions
take the form of the following first-order quasilinear system of PDEs
\begin{equation}
u^1_t = u^2_x,\quad u^2_t = u^3_x,\quad u^3_t = ((u^2)^2 -
u^1u^3)_x.\label{eq:37}
\end{equation}
We will call such a system a \emph{first-order WDVV system}.
It is of the general conservative form
\begin{equation}\label{eq:3}
  u^i_t = (V^i(\mathbf{u}))_x = \pd{V^i}{u^j}u^j_x,
\end{equation}
where $u^i=u^i(t,x)$ are field variables, $i=1$, \dots, $n$, $n$ is the number
of $x$-derivative variables, and $\mathbf{u}=(u^1,\dots,u^n)$.  We observe that
it is possible to reconstruct a solution $F$ (up to second-degree polynomials
in the independent variables) from a solution of the above system of
conservation laws (see Proposition~\ref{pro:sol-wdvv}).

For such systems it is possible to look for a bi-Hamiltonian
formalism with respect to two \emph{compatible} Hamiltonian operators~$A_{1}$
and~$A_{2}$,
\begin{equation*}
  u_{t}^{i}=A_{1}^{ij}\fd{H_2}{u^j}=A_{2}^{ij}\fd{H_{1}}{u^j}.
\end{equation*}
We recall that an operator~$A$ is called Hamiltonian if and only
if the operation
\begin{equation*}
  \{F,G\}_{A} = \int\fd{F}{u^i}A^{ij} \fd{G}{u^j}\,\mathrm dx
\end{equation*}
is a Poisson bracket, that is, if it is skew-symmetric and satisfies the Jacobi
identity (see e.g.~\cite{KrasilshchikVinogradov:SCLDEqMP,
  Dorfman:DSInNEvEq,GelfandDorfman:SBHOp,
  KVV17,magri08:_hamil_poiss,Olver:ApLGDEq}).  Compatibility of two Hamiltonian
operators~$A_1$ and~$A_2$ means that any linear combination
$\lambda A_1 + \mu A_2$ yields another Poisson bracket.

More generally, it was proved in~\cite{vasicek21:_wdvv_hamil} that in dimension
$N=3$ the WDVV system admits a bi-Hamiltonian formalism for any choice of~$\eta$.
In all cases, one of the Hamiltonian operators is a homogeneous first-order
operator of Ferapontov type \cite{F95:_nl_ho}
\begin{gather}\label{FerHamOpGen}
  A_1^{ij} = g^{ij}\mathrm D_x{} + \Gamma^{ij}_{k}u^k_x
  + c^{\alpha\beta} w^i_{\alpha q}u^q_x\mathrm D_x^{-1} w^j_{\beta p}u^p_x,
\end{gather}
where $c^{\alpha\beta}$ is a nondegenerate constant symmetric matrix and all
other coefficients are functions of $(u^i)$ (which correspond to the
$t^p$-derivative variables).  The other Hamiltonian operator is a homogeneous
third-order operator of Dubrovin--Novikov type \cite{DubrovinNovikov:PBHT} in
the canonical Doyle--Pot\"emin form~\cite{balandin01:_poiss,
  doyle93:_differ_poiss,potemin91:PhDt,potemin97:_poiss},
\begin{equation}
  \label{DPHamOp}
  A_2^{ij} = \mathrm D_x\circ(f^{ij}\mathrm D_x + c^{ij}_k u^k_x)\circ\mathrm
  D_x.
\end{equation}

Such a pair is not limited exclusively to WDVV equations.  Indeed,
in~\cite{OpanasenkoVitolo2024} we considered such ``WDVV-type'' systems
and classified them in low dimensions.

When $N=4$ and $\eta=\eta^{(1)}$ it was proved~\cite{ferapontov96:_hamil} that
the compatibility conditions for the new variables (third-order derivatives
of~$f$) can be written, using the WDVV equations, as a pair of \emph{commuting}
hydrodynamic type systems in a conservative form
\begin{alignat*}{2}
  \begin{aligned} & \begin{cases}
      u^1_y = u^2_x,\\
      u^2_y = u^3_x,\\
      u^3_y = \left(\dfrac{2u^{5}+u^{2}u^{3}}{u^{1}}\right)_x,\\[1ex]
      u^4_y = u^5_x,\\[3ex]
      u^5_y = \left(\dfrac{u^{3}u^{4}+u^{6}}{u^{1}}\right)_x,\\[3ex]
      u^6_y = \left(\dfrac{2u^{4}u^{5}-u^{2}u^{6}}{u^{1}}\right)_x;
  \end{cases}
  \end{aligned}
  \qquad
  \begin{aligned}
    & \begin{cases}
      u^1_z =u^4_x,\\
      u^2_z =u^5_x,\\
      u^3_z =\left(\dfrac{u^{3}u^{4}+u^{6}}{u^{1}}\right)_x,\\[1ex]
      u^4_z =u^6_x,\\[3ex]
      u^5_z =\left(\dfrac{2u^{4}u^{5}-u^{2}u^{6}}{u^{1}}\right)_x,\\[3ex]
        u^6_z =\left((u^{5})^{2}-u^{3}u^{6}+\dfrac{u^{3}(u^{4})^{2}
        + u^{4}u^{6}-2u^{2}u^{4}u^{5}+(u^{2})^{2}u^{6}}{u^{1}}\right)_x.
  \end{cases}
  \end{aligned}
\end{alignat*}
Again, the equivalence of the above systems with the WDVV system holds on
the open dense subset of the space of dependent variables where the denominator
$u^1$ does not vanish.

The Authors of~\cite{ferapontov96:_hamil} also proved that the above systems
admit the same \emph{local} ($c^{\alpha\beta}=0$) Ferapontov-type Hamiltonian
operator, of course, with two distinct Hamiltonian densities.
It was later found in~\cite{PV15} that the above two systems admit the same
third-order operator of the type~\eqref{DPHamOp}, and that the latter operator
is \emph{compatible} with the former operator.

Hamiltonian operators were also found in the cases $N=4$,
$\eta=\eta^{(2)}$ and $N=5$, $\eta=\eta^{(1)}$, $\eta=\eta^{(2)}$
\cite{lorenzoni25:_compat_hamil,vasicek21:_wdvv_hamil,vasicek22:_wdvv}.

It should be remarked that, since the above systems of conservation laws (one
in dimension $N=3$ and two in dimension $N=4$), admit a third-order Hamiltonian
operator of the form~\eqref{DPHamOp}, they admit an interpretation as
distinguished projective varieties in the Pl\"ucker embedding of $\mathbb{P}^N$
\cite{FPV17:_system_cl}. Namely, following a construction by
\cite{agafonov96:_system} (see also
\cite{agafonov99:_theor,agafonov01:_system_templ}), the space $\mathbb{P}^N$ is
the projective space whose affine coordinates are $(u^i)$, and the varieties
are linear line congruences. These ideas will be developed in
Section~\ref{sec:wdvv-equations-as}.

\textbf{Problem:} is it always true that, by introducing new variables, we can
express WDVV equations as a finite number of \emph{commuting} quasilinear
systems of first-order conservation laws?  If the above is true, is it always
true that the commuting quasilinear systems of first-order conservation laws
will have \emph{the same} Hamiltonian operators? Can we reconstruct solutions
of WDVV equations from joint solutions of the commuting quasilinear systems?

\subsection{Main results}

We have been able to prove the following results.

\paragraph{Hamiltonian formalism.} The WDVV equations contain subsystems that
can be rewritten, after a choice of a distinguished independent variable
$t^p$, as $N-2$ quasilinear first-order systems of conservation laws, each of
which is Hamiltonian with respect to the same homogeneous third-order
Hamiltonian operator in canonical form~\eqref{DPHamOp}, see
Theorem~\ref{theor:HamTheorem}. The steps to achieve this result, which holds
for generic $N$ and $\eta$, are
\begin{itemize}
\item the Structure Theorem~\ref{theor:WDVVLinComplexForm}, where we observe
  that WDVV equations contain subsystems that can be identified with linear
  line congruences in suitable Pl\"ucker embeddings;
\item the Metric Theorem~\ref{theor:metric-theorem}, where we complete the
  proof of the existence and uniqueness of a Hamiltonian operator as above.
\end{itemize}
It is remarkable that the Hamiltonian operator is one and the same for all
these $N-2$ subsystems (see the Hamiltonian Theorem~\ref{theor:HamTheorem}).

We observe that a third-order Hamiltonian operator as above is uniquely
determined by its leading coefficient matrix (in the nondegenerate case
$\det(f^{ij})\neq 0$), which is a Monge metric (see~\cite{FPV14,FPV16}). Such a
metric determines a quadratic line complex, which is a variety of lines in the
Pl\"ucker embedding of the projective space whose affine coordinates are the
field variables. In other words, the choice of an independent variable $t^p$
determines a quadratic line complex, which `sits' in WDVV equations. We expect
this object to play a role beyond the algebraic structure considered here.

We could not prove that all first-order WDVV systems can be endowed with a
first-order Hamiltonian operator. However, we proved by direct computation
(results in \cite{lorenzoni25:_compat_hamil}) that in the dimensions $N=4$ and
$N=5$ the first-order WDVV systems associated with $\eta=\eta^{(2)}$, and
obtained after the choice $t^p=t^2$, admit a Ferapontov-type first-order
nonlocal Hamiltonian operator of the form~\eqref{FerHamOpGen}. Such operator is
compatible with the third-order operator, hence first-order WDVV systems turn
out to be bi-Hamiltonian of WDVV-type. See the discussion in
Section~\ref{sec:wdvv4_HamOp}.

It is natural to \emph{conjecture} that all first-order WDVV systems admit a
first-order operator of Ferapontov type that is compatible with the
third-order operator found in this paper for all such systems.

\paragraph{Reducibility, compatibility and commutativity.}
Compatibility of WDVV equations is still an open problem. We observed that such
conditions are at the heart of the commutativity of the above systems of
conservation laws. We were able to prove the following results.
\begin{itemize}
\item The WDVV equations, together with the quasihomogeneity assumption, can be
  reduced, for arbitrary $N$ and $\eta$, as follows. Having chosen an
  independent variable $t^p$, the WDVV equations can be solved for all
  $t^p$-free third-order derivatives, and are equivalent to an orthonomic
  system of the form
  \begin{equation*}
    f_{ijk}=G^{ijk}(f_{plm}),\qquad i,j,k\in \{2,\ldots,N\}\setminus\{p\}.
  \end{equation*}
  That is the content of the Reduction Theorem~\ref{th:reduc-wdvv-equat}.
\item For arbitrary $N$ and $\eta$, the above system is compatible (or passive)
  if and only if the $N-2$ Hamiltonian first-order systems of quasilinear
  equations determined by the choice of~$t^p$ commute
  (Theorem~\ref{theor:Commutativity}).  Compatibility, or passivity, means that
  the system does not contain any hidden integrability conditions.
\item Again in the general situation of arbitrary $N$ and $\eta$, if passivity
  holds, then we are able to integrate any joint solution of the $N-2$
  Hamiltonian systems to a solution of the WDVV equations up to second degree
  polynomials of the independent variables
  (Proposition~\ref{pro:reduc-wdvv-equat}).
\item In low dimensional cases ($N=4$ and arbitrary $\eta$, $N=5$ and
  $\eta =\eta^{(2)}$) we were able to prove by direct symbolic calculation that
  commutativity holds (Theorem~\ref{prop:reduc-wdvv}). That leads us to
  conjecture that commutativity holds in general.
\end{itemize}

\subsection{Computing}

The calculations have been done by means of the computer algebra systems Maple
and Reduce (with its package CDE, see~\cite{KVV17}). More precisely, our
calculations concern two main topics:
\begin{itemize}
\item we prove that it is possible to reduce the number of algebraically
  independent WDVV equations and formulate the residual equations as $N-2$
  commuting quasilinear first-order systems of conservation laws, in the
  following cases. In dimension $N=4$, we computed with arbitrary $\eta$
  (Maple), and Dubrovin's canonical form $\eta^{(2)}$ (Reduce), in dimension
  $N=5$ we computed with Dubrovin's canonical form $\eta^{(2)}$ (both Maple and
  Reduce). Note that the calculation with $N=4$ and arbitrary~$\eta$ takes 60GB
  RAM and about 12 hours running on a compute server;
\item we prove the base cases of the Metric Theorem ($N\leq 9$). We also add
  some computations that support the proofs of the Lemmas that are involved in
  it.
\end{itemize}

The calculations are publicly available at a GitLab repository of the Istituto
Nazionale di Fisica Nucleare (INFN) \cite{opavit26:_maple_reduc_wdvv}.

\section{Preliminaries}

\subsection{The metric \texorpdfstring{$\eta$}{eta}
  and related constructions}

We consider the metric $\eta$, which is the crucial data in the framework of
WDVV equations, as the nondegenerate symmetric bilinear form
\begin{equation*}
  \eta\colon \rn\times \rn \to \mathbb{R}.
\end{equation*}
Let us denote by $\bar{\eta}$ the inverse metric on $(\rn)^*$.
If $\{e_i\mid i=1,\ldots,N\}$ is a basis of $\rn$ and $\{e^i\mid i=1,\ldots,N\}$
is the dual basis of $(\rn)^*$, then we set $\eta_{ij}:=\eta(e_i,e_j)$.
The matrix $(\eta_{ij})$ is invertible; its inverse is denoted by $(\eta^{ij})$,
and it is the matrix of $\bar{\eta}$.
We can regard the metric $\eta$ as a nondegenerate, symmetric tensor
$\eta\in (\rn)^*\otimes (\rn)^*$; its inverse is identified with the symmetric
nondegenerate tensor $\bar{\eta}\in \rn\otimes \rn$.
The metric $\eta$ defines a metric $\eta^r_s$ on the space of $r$-contravariant
and $s$-covariant tensors
$\otimes^r_s \rn=\otimes^r \rn\otimes\otimes^s (\rn)^*$ as follows. On
decomposable elements
\begin{equation*}
  t=v_1\otimes\cdots\otimes v_r\otimes\alpha^1\otimes\cdots\otimes\alpha^s,
  \quad
 u=w_1\otimes\cdots\otimes w_r\otimes\beta^1\otimes\cdots\otimes\beta^s
\end{equation*}
we define
\begin{equation*}
  \eta^r_s(t,u)=\left(\prod_{i=1}^r\eta(v_i,w_i)\right)\left(\prod_{j=1}^s\bar{\eta}(\alpha^j,\beta^j)\right),
\end{equation*}
and uniquely extend the definition to arbitrary tensors by
bilinearity. Nondegeneracy is again clear. Recalling that a basis of
$\otimes^r_s \rn$ is given by all elements of the form
$e_{i_1}\otimes\cdots\otimes e_{i_r}\otimes e^{a_1}\otimes\cdots\otimes
e^{a_s}$, the coordinate expression of $\eta^r_s$ is
\begin{multline*}
  \eta^r_s(e_{i_1}\otimes\cdots\otimes e_{i_r}\otimes e^{a_1}\otimes\cdots\otimes
  e^{a_s},
  e_{j_1}\otimes\cdots\otimes e_{j_r}\otimes e^{b_1}\otimes\cdots\otimes e^{b_s})
= \eta_{i_1j_1}\cdots\eta_{i_rj_r}\eta^{a_1b_1}\cdots\eta^{a_sb_s}.
\end{multline*}

The above metric can be restricted to the exterior algebra of $r$-vectors
$\wedge^r \rn\subset\otimes^r_0 \rn = \otimes^r \rn$. Recalling that a basis of
$\wedge^r \rn$ is provided by
$\{e_{i_1}\wedge\cdots\wedge e_{i_r}\mid 1\leq i_1<\cdots<i_r\leq N\}$ and that
\begin{equation*}
  e_{i_1}\wedge\cdots\wedge e_{i_r} =
  \frac{1}{r!}\sum_{\sigma\in S_r}|\sigma| e_{\sigma(i_1)}\otimes\cdots
  \otimes e_{\sigma(i_r)},
\end{equation*}
where $S_r$ is the set of permutation of $\{1,\ldots,r\}$,
we have the coordinate expression
\begin{equation*}
  \eta^r_0(e_{i_1}\wedge\cdots\wedge e_{i_r},e_{j_1}\wedge\cdots\wedge e_{j_r})
   = \frac{1}{r!}\eta_{i_1\cdots i_r,j_1\cdots j_r}
\end{equation*}
where by $\eta_{i_1\cdots i_r,j_1\cdots j_r}$ we indicate the \emph{retainer}
minor~\cite[p.~18]{vein99:_deter_their_applic_mathem_physic} of $\eta=(\eta_{ij})$,
made by selecting the rows $i_1$, \ldots, $i_r$ and the columns $j_1$, \ldots, $j_r$
and then taking the determinant of the resulting square submatrix of order~$r$.
In particular, when $r=2$ we have
\begin{equation*}
  \eta^2_0(e_{i_1}\wedge e_{i_2},e_{j_1}\wedge e_{j_2}) = \eta_{i_1i_2,j_1j_2}
:= \frac{1}{2}(\eta_{i_1j_1}\eta_{i_2j_2} - \eta_{i_1j_2}\eta_{i_2j_1}).
\end{equation*}
We can repeat the entire construction for $s$-forms
$\wedge^s (\rn)^*\subset\otimes^0_s \rn = \otimes^s (\rn)^*$, and obtain
the inverse metric $\eta^0_s$. In particular,
\begin{equation*}
  \eta^0_2(e^{i_1}\wedge e^{i_2},e^{j_1}\wedge e^{j_2}) = \eta^{i_1i_2,j_1j_2}
   = \frac{1}{2}(\eta^{i_1j_1}\eta^{i_2j_2} - \eta^{i_1j_2}\eta^{i_2j_1}).
\end{equation*}
In what follows, we will rescale both $\eta^2_0$ and $\eta_2^0$ by a factor $2$
in order to have simpler coordinate expressions.

We can regard a metric on $\wedge^2(\rn)^*$ as a symmetric tensor
\begin{equation*}
  \phi\in S^2(\wedge^2\rn)\subset (\wedge^2\rn)\otimes(\wedge^2\rn),
\end{equation*}
where we have denoted by $S^2(\wedge^2\rn)$ the subspace of symmetric
tensors. We observe that there is a natural linear map
\begin{equation*}
  \label{eq:23}
  C\colon S^2(\wedge^2\rn) \to\wedge^4\rn,\quad X\otimes Y \mapsto X\wedge Y.
\end{equation*}
The coordinate expression of the above map~$C$ is obtained as follows.
Let $\phi\in S^2(\wedge^2\rn)$ be of the form
\begin{equation*}
  \phi^{i_1j_1,i_2j_2}e_{i_1}\wedge e_{j_1}\otimes e_{i_2}\wedge e_{j_2},
\end{equation*}
where the sum is extended to all indices fulfilling the conditions
$i_1<j_1$ and $i_2<j_2$. Then, the image of the map~$C$ is
\begin{equation*}
  \phi^{i_1j_1,i_2j_2}e_{i_1}\wedge e_{j_1}\wedge e_{i_2}\wedge e_{j_2},
\end{equation*}
where $i_1<j_1$ and $i_2<j_2$. In order to express the above $4$-form with
respect to the natural basis of~$\wedge^4\rn$ we collect similar terms and sum
indices under the condition $a<b<c<d$, obtaining
\begin{equation*}
  2(\phi^{ab,cd} - \phi^{ac,bd} + \phi^{ad,bc})
  e_a\wedge e_b\wedge e_c\wedge e_d,
\end{equation*}
with the help of the symmetry $\phi^{ab,cd} = \phi^{cd,ab}$.

\begin{definition}
  We say that a symmetric tensor $\phi\in S^2(\wedge^2\rn)$ is in Clebsch
  canonical form if $\phi\in\ker C$.
\end{definition}
The above definition is a generalization of Clebsch canonical form in~$\mathbb{P}^3$
(see~\cite{FPV16,jessop03} for more details). We have the
following result.
\begin{lemma}
  $\eta^0_2\in\ker C$.
\end{lemma}
\begin{proof}
  In coordinates, the equations of the kernel of the natural map are just the
  Pl\"ucker quadrics, and the components of $\eta^0_2$ are $2\times 2$ minors,
  which obviously fulfill the equations.
\end{proof}

\subsection{Hamiltonian structures of WDVV-type}
\label{sec:hamilt-struct-wdvv}

In this section we will introduce the properties that make the operators
$A_1$~\eqref{FerHamOpGen} and $A_2$~\eqref{DPHamOp} Hamiltonian. See
\cite{OpanasenkoVitolo2024} for more details, especially about the
computational aspects. We stress that we consider the nondegenerate case: we
require that the conditions $\det(g^{ij})\neq 0$, $\det(f^{ij})\neq 0$ hold on
a dense open subset of the space of the dependent variables.

We stress that the results of this paper concern mostly third-order Hamiltonian
operators, although we plan to fully address first-order Hamiltonian operators
in the near future.

\paragraph{Third-order operator.}
\label{sec:third-order-operator}

The operator~$A_2$~\eqref{DPHamOp} is Hamiltonian if and only if
\begin{gather}\label{eq:19}
  c_{skm}=\frac{1}{3}(f_{sm,k} - f_{sk,m}),
  \\
  f_{mk,s} + f_{ks,m} + f_{ms,k}=0, \label{eq:20}
  \\
  c_{msk,l}= - f^{pq}c_{pml}c_{qsk}, \label{eq:21}
\end{gather}
where $c_{ijk}=f_{iq}f_{jp}c_{k}^{pq}$, so that the leading coefficient
completely determines the operator (in the nondegenerate case,
see~\cite{FPV14}).

It was found out in \cite{FPV14} that equation~\eqref{eq:20} is equivalent to
the fact that $f_{ij}$ is the Monge form, or Monge metric, of a quadratic line
complex, which is a projective variety in the Pl\"ucker embedding
of~$\mathbb{P}^n$. Here $\mathbb{P}^n$ is the projective space whose affine
chart is given by the variables $(u^i)$.

A complete classification of operators in the form~\eqref{DPHamOp}, up to the
projective invariance with respect to reciprocal transformations, is given
in~\cite{FPV14,FPV16} for a number of components $n\leq 4$.
It was proved~\cite{balandin01:_poiss,FPV14} that the Monge metric of a
third-order Hamiltonian operator can be factorised as follows:
\begin{equation*}
  \label{eq:30}
  f = \phi_{\alpha\beta}\psi^\alpha_i \psi^\beta_j\mathrm du^i\otimes\mathrm du^j.
\end{equation*}
Here, $\psi^\alpha_i = \psi^\alpha_{ik}u^k + \omega^\alpha_i$ and
$\psi^\alpha_{ji}= - \psi^\alpha_{ij}$, where the indices $\alpha$, $i$, $j$
run from $1$ to $n$. The indexed objects $\phi_{\alpha\beta}$,
$\psi^\alpha_{ij}$, $\omega^\alpha_i$ are constants. Note that the matrices
$\phi_{\alpha\beta}$, $\psi^\alpha_i$ must be nondegenerate on a dense open
subset of the space of dependent variables, as $f$ is
nondegenerate in the same sense. The algebraic variety $\det (\psi^\alpha_i) =
0$ has a geometric interpretation: it is the singular surface of the quadratic
line complex identified by $f$ (see \cite{FPV16} for more details).

We introduce the additional coordinate $u^{n+1}$ and rewrite
$\psi^\alpha_i\mathrm du^i$ as follows
\begin{equation}
  \label{eq:31}
  \frac{1}{2}\psi^\alpha_{ij}(u^j\mathrm du^i - u^i\mathrm du^j) +
  \frac{1}{2}\psi^\alpha_{(n+1)i}(u^{n+1}\mathrm du^i - u^i\mathrm du^{n+1}),
\end{equation}
where $\psi^\alpha_{(n+1)i} = - \psi^\alpha_{i(n+1)}=\omega^{\alpha}_i$. Note
that we can get back $\psi^\alpha_i\mathrm du^i$ by means of a projection to
affine coordinates $u^{n+1}=1$, $\mathrm du^{n+1}=0$. Let us now introduce the
indices $a$, $b$ running from $1$ to $n+1$. Under the action of linear
transformations of $\mathbb{R}^{n+1}$, the one-forms
\begin{equation*}
  \frac{1}{2}(u^a\mathrm du^b - u^b\mathrm du^a)
\end{equation*}
transform like two-forms $e^a\wedge e^b$, where $\{e^a\mid a=1,\ldots,n+1\}$ is
a basis of $(\mathbb{R}^{n+1})^*$. So, we can rewrite
$\psi^\alpha_i\mathrm du^i$ as a family of two-forms
\begin{equation}\label{eq:38}
  A^\alpha = \psi^\alpha_{ab}e^b\wedge e^a,\qquad \alpha = 1,\ldots,n.
\end{equation}
It was proved in \cite{FPV16} that a third-order homogeneous Hamiltonian
operator in the Doyle--Pot\"emin form~\eqref{DPHamOp} can be identified with a
linear $n$-dimensional subspace
\begin{equation*}
  A = \Span(\{A^\alpha\mid \alpha=1,\ldots,n\})
  \subset \wedge^2\mathbb{R}^{n+1}
\end{equation*}
and a metric $\phi$ on $A$, where the metric $\phi$ is requested to lie in the
kernel of the natural restriction
\begin{equation*}
  C|_A\colon S^2(A) \to A\wedge A,
\end{equation*}
or $\phi\in\ker(C|_A)=\ker C\cap S^2(A)$. In coordinates, the following
equations must hold:
\begin{subequations}\label{eq:48}
\begin{align}
  & \phi _{\beta \gamma }(\psi _{is}^{\beta }\psi _{jk}^{\gamma }+\psi
    _{js}^{\beta }\psi _{ki}^{\gamma }+\psi _{ks}^{\beta }\psi _{ij}^{\gamma
    })=0, \label{ab}%
  \\
  & \phi _{\beta \gamma }(\omega _{i}^{\beta }\psi _{jk}^{\gamma
    }+\omega _{j}^{\beta }\psi _{ki}^{\gamma }+\omega _{k}^{\beta }\psi
    _{ij}^{\gamma })=0. \label{zac}%
\end{align}
\end{subequations}

The conditions under which systems of the type~\eqref{eq:3} are Hamiltonian
with respect to $A_2$ are best expressed through the vector function
$Z^\alpha = \psi^\alpha_i V^i$.
Indeed, it was proved in \cite{FPV17:_system_cl} that any first-order system of
conservation laws of the type~\eqref{eq:3} is Hamiltonian with respect to a
third-order operator $A_2$ of the type~\eqref{DPHamOp} if and only if
$Z^\alpha$ are linear functions, $Z^\alpha = \theta^\alpha_i u^i + \xi^i$, and
their coefficients fulfill the algebraic system
\begin{subequations}\label{eq:53}
\begin{align}\label{eq:54}
  & \phi_{\beta\gamma}[\psi_{ij}^{\beta}\theta_{k}^{\gamma}+\psi_{jk}^{\beta}
    \theta_{i}^{\gamma}+\psi_{ki}^{\beta}\theta_{j}^{\gamma}]=0,
  \\ \label{eq:55}
  & \phi_{\beta\gamma}[\psi_{ik}^{\beta}\xi^{\gamma}+\omega_{k}^{\beta}
    \eta_{i}^{\gamma}-\omega_{i}^{\beta}\eta_{k}^{\gamma}]=0.
\end{align}
\end{subequations}
The above equations~\eqref{eq:48}, \eqref{eq:53} can be assembled together
using the space $(\mathbb{R}^{n+2})^*$, which extends $(\mathbb{R}^{n+1})^*$
with the basis $\{e^1,\ldots,e^{n+2}\}$,
and the extended family of $n$ two-forms
\begin{equation}
  \label{eq:28}
  \bar{A}^{\alpha} = \psi^\alpha_{ab} e^b \wedge e^a
  - \theta ^\alpha_i e^i\wedge e^{n+2} - \xi^\alpha e^{n+1}\wedge e^{n+2},
\end{equation}
where indices run as specified above. Then,~\eqref{eq:48} and \eqref{eq:53} are
condensed into $C|_{\bar{A}}(\phi)=0$, where $C|_{\bar{A}}\colon S^2(\bar{A})
\to \bar{A}\wedge\bar{A}$.

A projective geometric interpretation is obtained as follows.  Expanding the
relation $\psi^\alpha_i V^i = Z^\alpha$ we obtain the equation
\begin{equation}
  \label{eq:9}
  \frac{1}{2}\psi^\alpha_{hk}(u^kV^h - u^hV^k) + \omega^\alpha_h V^h -
  \theta^\alpha_k u^k - \xi^\alpha = 0.
\end{equation}

On the other hand, following the approach of
\cite{agafonov96:_system,agafonov99:_theor,agafonov01:_system_templ}, we
associate a \emph{congruence of lines} to any system of the type~\eqref{eq:3},
\emph{i.e.} an $n$-parameter family of lines
\begin{equation*}
  y^i = u^iy^{n+1} + V^i y^{n+2}
\end{equation*}
in a projective space $\mathbb{P}^{n+1}$ with homogeneous coordinates
$[y^1,\ldots, y^{n+2}]$. More precisely, the equation determines an
$n$-parameter family of lines passing through the points
$y^i=u^i, \ y^{n+1}=1, \ y^{n+2}=0$ and $y^i=V^i, \ y^{n+1}=0, \
y^{n+2}=1$. The Pl\"ucker coordinates of the above family of lines in the
Pl\"ucker embedding of $\mathbb{P}^{n+1}$ are the $2\times 2$ minors of the
matrix
\begin{equation*}
  \begin{pmatrix}
    u^i & \dots & u^n & 1 & 0\\
    V^i & \dots & V^n & 0 & 1
  \end{pmatrix}.
\end{equation*}
So, if a first-order system of conservation laws is Hamiltonian with respect to
a third-order operator as above, then the associated congruence is
linear:~\eqref{eq:9} is a linear expression of the Pl\"ucker's coordinates.

The properties of linear line congruences in connection with systems of
first-order conservation laws of PDEs have been widely investigated
\cite{agafonov98:_linear,agafonov96:_system,agafonov99:_theor,agafonov01:_system_templ}.
Here, we recall that a system of first-order
conservation laws that admits a Hamiltonian structure of the
type~\eqref{DPHamOp} (and hence whose associated line congruence is linear) is
linearly degenerate and in the Temple class~\cite{FPV17:_system_cl}.

The relation \eqref{eq:9} will be of crucial importance for our results.

\paragraph{First-order operator.}
\label{sec:first-order-operator}

The operator~$A_1$~\eqref{FerHamOpGen} is Hamiltonian if and only if the
following conditions hold:
\begin{itemize}
\item $(g^{ij})$ is a symmetric matrix; its inverse $(g_{ij})$ can be
  interpreted as a metric, and the Christoffel symbols of its Levi-Civita
  connection are $\Gamma^i_{jk}= - g_{jp}\Gamma^{pi}_k$;
\item $w^i_{\alpha j}$ are symmetric endomorphisms with respect to $(g_{ij})$,
  fulfilling the additional conditions
  \begin{subequations}\label{eq:12}
  \begin{gather}
    \label{eq:65} \nabla_k w^i_{\alpha j} = \nabla_j w^i_{\alpha k},
    \\
    [w_\alpha,w_\beta] = 0,\label{eq:17}
    \\
    \label{eq:63} R^{ij}_{hl} =
    c^{\alpha\beta}\Big( w^i_{\alpha l}w^j_{\beta h}
    - w^i_{\alpha h}w^j_{\beta l}\Big);
  \end{gather}
\end{subequations}
\item $c^{\alpha\beta}$ is a symmetric nondegenerate matrix.
\end{itemize}
In the above equations, $\nabla$ is the Levi-Civita connection of $(g_{ij})$,
and $R^{ij}_{hl}$ is its Riemannian curvature; see~\cite{OpanasenkoVitolo2024}
for more details. The condition~\eqref{eq:17} is the vanishing of the Jacobi
bracket~\cite{KrasilshchikVinogradov:SCLDEqMP} of generalized vector
fields $w_\alpha=w_{\alpha\,j}^iu^j_x\pd{}{u^i}$.  If
$w_{\alpha\,j}^iu^j_x = (w_\alpha^i)_x$, then the condition reduces to the
commutativity of the matrix product
$w^i_{\alpha k} w^k_{\beta j} = w^i_{\beta k} w^k_{\alpha j}$.

WDVV-type systems are of the form~\eqref{eq:3}. The conditions under which such
a system is Hamiltonian with respect to an operator $A_1$ of the above
type~\cite{tsarev85:_poiss_hamil} (see also~\cite{vergallo20:_homog_hamil}) are
\begin{gather*}
g^{ik}V^j_k = g^{jk}V^i_k,\quad
\nabla^iV^j_k = \nabla^j V^i_k,
\end{gather*}
where $V^i_k = \pd{V^i}{u^k}$, and the vector field $V=(V^i)_{x}\pd{}{u^i}$
commutes with the vector fields $w_\alpha$ with respect to the Jacobi bracket.

\section{First-order WDVV systems}
\label{sec:invar-wdvv-equat}

As we have mentioned in Introduction, the WDVV equations~\eqref{eq:5} are an
overdetermined nonlinear system of PDEs in the unknown function
$f=f(t^2,\ldots,t^N)$. In low dimensions and for particular values of $\eta$ it
is known to be represented as systems of conservation laws by introducing new
dependent variables (see below for references).

Here, we would like to write down the transformation of WDVV equations into
first-order WDVV systems for an arbitrary choice of the matrix $\eta$ and an
arbitrary value of the dimension~$N$. In order to do that, we will clarify
some aspects that do not seem to have been dealt with in detail elsewhere.

Our framework is that of the Geometry of PDEs, which is based on jet spaces.

\subsection{WDVV equations as a submanifold of a jet space}
\label{sec:wdvv-systems}

In this Subsection, we will give a geometric description of WDVV equations and
the first-order quasilinear systems of conservation laws that we can construct
from them. The natural framework is that of jet spaces. We will use results in
\cite{MaMo83}, see also~\cite{Saunders:GJB}.

As we consider the local theory of WDVV equations, we use the trivial bundle
$\pi\colon E=M\times F\to M$, where $M\subset \mathbb{R}^{N-1}$ and
$F\subset \mathbb{R}$ are two open subsets; a fibered chart on~$E$ has the form
$(t^\lambda,f)$, $2\leq \lambda\leq N$.  The WDVV equation is a submanifold
$\mathcal{E}\subset J_3E$ that we want to regard as a family of first-order
quasilinear systems of PDEs. To this aim, we consider the first jet
prolongation~$\mathcal{E}^{(1)}$ of the WDVV system (see
\cite{KrasilshchikVinogradov:SCLDEqMP,Olver:ApLGDEq}), which in coordinates is
defined as
\begin{equation}
  \label{eq:4}
  \mathcal{E}^{(1)} = \{\partial_\lambda S_{\mu\nu\rho\sigma} = 0\mid
  \lambda,\mu,\nu,\rho,\sigma =1,\ldots N\}\subset J_4E.
\end{equation}
The number of third-order derivative coordinates $f_\tau$ in $J_3E$, where
$\tau$ is a multiindex $\tau\in\mathbb{N}^{N-1}$, $|\tau|=3$, is
$M=\binom{N+1}{3}$. Note that we number the independent variables starting from
$2$, so that $\tau = (\tau_2,\ldots,\tau_N)$.

By interpreting $J_3E$ as a fibred manifold over $M$ we can introduce a
subbundle of its first jet space $\hat{J}_4E\subset J_1J_3E$, which is called
the \emph{sesquiholonomic $4$-th jet space}, characterized as the kernel of the
Spencer operator on iterated jet spaces \cite{MaMo83} (see
also~\cite{Saunders:GJB}). More precisely, $\hat{J}_4E$ is an affine subbundle
whose associated vector bundle is
$T^*M\otimes (\odot^3T^*M)) \otimes_{J_3E} \ker T\pi$. Here, $\odot$ is the
symmetric tensor product.  We have the affine direct sum decomposition
\begin{equation}
  \label{eq:27}
  \hat{J}_{4}E = J_{4}E \oplus_{J_3E} S_4,
\end{equation}
where $S_4$ is the vector bundle
$S_4=(T^*M\wedge_M (\odot^3T^*M)) \otimes_{J_3E} \ker T\pi$.
It turns out that $J_4E$ is the subbundle of $\hat{J}_4E$ that is
determined by the vanishing of the projection onto the right-hand summand of the
above splitting.

In coordinates, if $(f,f_{\tau_1},f_{\tau_2},f_{\tau_3},f_{\tau_3',\lambda})$ are
derivative coordinates of $\hat{J}_{4}E$ where $\|\tau_i\|=i$, then $J_4E$ is
the solution space of equations
\begin{equation}
  \label{eq:29}
  f_{\tau_3,\lambda} - f_{\tau_3+\lambda} = 0
\end{equation}
where $\tau_3+\lambda$ is the multiindex obtained by $\tau_3$ by summing $1$ to
the $\lambda$-th component of $\tau_3$. The main idea here is to
restrict~\eqref{eq:29} to $\mathcal{E}^{(1)}$ and represent that restriction
with a set of necessary and sufficient conditions. We can
rewrite~\eqref{eq:29} as
\begin{equation}
\label{eq:33}
  f_{\tau_2+\mu,\lambda} - f_{\tau_2+\lambda,\mu} = 0,\quad\text{where}\quad
  \mu\leq\lambda.
\end{equation}
The number of the above equations where $\mu<\lambda$ is the number of all
multiindices $\tau_2$, which is ${N(N-1)/2}$, multiplied by the number of all
pairs $(\mu,\lambda)$, where $\mu<\lambda$, which is ${(N-1)(N-2)/2}$.  In other
words, the conditions that have to be verified on coordinates in $\hat{J}_4E$
in order to identify a subbundle which is diffeomorphic to $J_4E$ reduce to
only~\eqref{eq:33} with $\mu<\lambda$.

It is of fundamental importance, for our induction proofs in next sections, to
introduce a ranking (in the sense of \cite{M09}) for derivatives of the
function~$f$ and for equations in the
system~$\mathcal S_{\alpha\beta\gamma\delta}$. Let $\sigma\in\mathbb{N}^{N-1}$.
We introduce the following linear ranking for derivatives
\begin{displaymath}
  f_\tau \ll f_\sigma \quad\Leftrightarrow\quad \exists i:
  \quad \text{$\tau_i<\sigma_i$ and $\tau_j=\sigma_j$ for all $j>i$.}
\end{displaymath}
Of particular interest to us is the ranking of second-order derivatives, since
we will use subsets of third-order derivatives $f_\sigma$ where, for a
distinguished index $i$ we have $\sigma_i> 0$. Let us momentarily drop the
multiindex notation, and denote a third-order derivative by~$f_{\lambda im}$,
where $\lambda\in\{2,\ldots,N\}$ is fixed and $i$, $j\in\{2,\ldots,N\}$, with
$i\leqslant m$. We introduce a ranking in the set of third-order derivatives
with a fixed~$\lambda$ by means of the following map.
\begin{definition}
  We define the \emph{rank map} to be the following map:
\begin{equation*}
  \rk\colon \{(m,i)\colon 2\leqslant i\leqslant m\leq N\} \to \{1,\ldots,n\},
  \quad\rk(m,i)= (m-1)(m-2)/2+i-1,
\end{equation*}
where $n=N(N-1)/2$.
We also set $\rk(i,m):=\rk(m,i)$.
\end{definition}
Intuitively, the ranking~$\rk(m,i)$ of the derivative~$f_{\lambda i m}$ is
equal to the sum of the ranking of the derivative
$f_{\lambda(m-1)(m-1)}$, ${\rk(m-1,m-1)}$, which is equal to the number of distinct derivatives
of order~2 of an $(m-2)$-variable function, $(m-1)(m-2)/2$, and a number of
derivatives ranked between them, $i-1$.  It is true since the derivatives that
are ranked lower than $f_{\lambda im}$ are $f_{\lambda jl}$, $j\leq l<m$, and
$f_{\lambda jm}$, $j<i$.  On another note, for every $\lambda$ the ranking of
the derivative $(f_{\lambda im})$ is the $(i-1,m-1)$th entry of the symmetric,
upper triangular matrix
\begin{displaymath}
  (f_{\lambda (i+1)(m+1)})
  =\begin{pmatrix}
\rk(2,2)  &  \rk(3,2)  &  \rk(4,2)  &  \rk(5,2) &\dots\\
    &  \rk(3,3)  &  \rk(4,3)  &  \rk(5,3) &\dots\\
    &               &  \rk(4,4)  &  \rk(5,4) &\dots\\
    &               &               &  \rk(5,5) &\dots
\end{pmatrix}
=
  \begin{pmatrix}
    1 & 2 & 4 & 7 & \dots
    \\
      & 3 & 5 & 8 & \dots
    \\
    &  & 6 & 9 & \dots
    \\
      &  &  & 10 & \dots
  \end{pmatrix}.
\end{displaymath}

\begin{lemma}
  The map $\rk$ is bijective.
\end{lemma}
\begin{proof}
  Indeed, let us assume that $(m,i)\neq (p,j)$. If $m=p$
it is easy to see that the values of $\rk$ are distinct; if, say, $p<m$ we
should prove that the equality
\begin{displaymath}
  \frac{1}{2}(m-1)(m-2)+i-1 = \frac{1}{2}(p-1)(p-2)+j-1
\end{displaymath}
cannot happen. We have
\begin{displaymath}
  \frac{1}{2}(m-1)(m-2) - \frac{1}{2}(p-1)(p-2) \geq \frac{1}{2}(m-1)(m-2) -
  \frac{1}{2}(m-2)(m-3) = m-2,
\end{displaymath}
while $j-i \leq m - 3$, so that $\rk(m,i)>\rk(p,j)$.
\end{proof}
In order to find a preimage $\rk^{-1}(l)$ one has to find a
maximal~$m\in\mathbb N$ such that $\frac12(m-1)(m-2)< l$ and set
$i:=l-\frac12(m-1)(m-2)+1$. In order to make the notation lighter, and in view
of the fact that~$\rk$ will be used in indices, we will write
$(m,i):=\rk(m,i)$.  We introduce the following
notation:
\begin{equation}
  \label{eq:50}
  u^{(m,i)}_\lambda:=f_{\lambda im}, \quad 2\leqslant i\leqslant m.
\end{equation}
We can indicate $(m,i)$ by its value $I\in\{1,\ldots,n\}$.  Then, the
system~\eqref{eq:33} can be rewritten as
\begin{equation}
  \label{eq:6}
  (u^I_{\mu})_\lambda = (u^I_{\lambda})_\mu,\quad\text{where}\quad
  \lambda,\mu\in\{2,\ldots,N\},\ \mu<\lambda.
\end{equation}
This change in notation is key to the geometric study of the restriction of
equations~\eqref{eq:6} to $\mathcal{E}^{(1)}$.

Let us choose one independent variable; without loss of generality, we can
choose $t^2$. Then, the system~\eqref{eq:6} contains $N-2$ systems of $n$
equations, with a $t^i$-derivative on the left-hand side ($i>2$) and the
$t^2$-derivative on the right-hand side, of the form
\begin{equation}\label{eq:8}
  (u^I_2)_3 = (u^I_3)_2,\quad \ldots,\quad
  (u^I_2)_N = (u^I_N)_2,\quad I=1,\ldots,n.
\end{equation}
Ideally, we would like that the above $N-2$ systems be `closed': their
right-hand sides should depend on $(u^I_2)$ only. That is what happens in the
systems in Subsection~\ref{sec:second-probl-find}.

However, that is not always the case. Indeed, we have two possibilities: in any
of the equations $(u^I_2)_k = (u^I_k)_2$ it can be that $u^I_k=f_{2im}$, with
$2\leqslant i\leqslant m$, or not. In the former case we say the equation to be
\emph{trivial}, in the latter case our objective will be to find the variable
$u^I_k$ as a function of $(u^I_2)$ through the WDVV equations.
\begin{proposition}\label{pro:trivialeq}
  In each of the $N-2$ systems~\eqref{eq:8} there are exactly $N-1$ trivial
  equations of the form $(u^I_2)_k = (u^I_k)_2$.
\end{proposition}
\begin{proof}
  Indeed, an equality of the type
  $(f_{t^it^jt^2})_{t^h} = (f_{t^it^jt^h})_{t^2}$ ($h\neq 2$) yields a trivial
  equation if and only if one of the indices $i$, $j$ is equal to $2$. In this
  case, the other index can assume all $N-1$ values.
\end{proof}
We note that the number of nontrivial
equations in any of the systems \eqref{eq:8} is $n-(N-1)=\frac12(N-1)(N-2)$.

We stress that if none of the indices $i$, $j$ is equal to $2$, the variable
$f_{t^it^jt^h}$ is not one of $(u^I_2)$, hence we will need to express it as a
function of $(u^I_2)$ using the WDVV equations. We would like to show that:
\begin{itemize}
\item equations~\eqref{eq:8} restrict to the first prolongation
  $\mathcal{E}^{(1)}\subset J_4E$ of the WDVV equations
  $\mathcal{E}\subset J_3E$;
\item all other equations in~\eqref{eq:6} vanish on account of
    the WDVV equations.
\end{itemize}
In order to achieve the goal, it is better to introduce meaningful
low-dimensional examples.

\subsection{A particular case: WDVV equations,
  \texorpdfstring{$\mathbf{N=3}$}{N=3}}

This case of WDVV equations is the most studied. The transformation of WDVV
into a quasilinear first-order system of conservation laws was first done
in \cite{mokhov95:_sympl_poiss} for the case $\eta=\eta^{(1)}$. The resulting
system was further studied in \cite{FGMN97,FM96:_equat_hamil}.  Other forms of
$\eta$ were considered in
\cite{kalayci98:_alter_hamil_wdvv,kalayci97:_bi_hamil_wdvv}. Here, we present
the first-order system for a generic $\eta$.

For simplicity, we denote $x=t^2$, $y=t^3$, $u^I = u^I_x$,
$v^I=u^I_y$. Using the ranking that we introduced in
Subsection~\ref{sec:wdvv-systems} we have
\begin{gather*}
  u^1=f_{xxx},\ u^2=f_{xxy},\ u^3=f_{xyy},
  \\
  v^1=f_{xxy}=u^2,\ v^2=f_{xyy}=u^3,\ v^3=f_{yyy}.
\end{gather*}
There is only one independent WDVV equation:~$S_{2233}=0$. Explicitly,
\begin{gather*}
  (f_{xyy}\eta^{22}+f_{yyy}\eta^{23}+\eta^{12}\eta_{33})f_{xxx}-f_{xxy}^2\eta^{22}
  +(-f_{xyy}\eta^{23}+f_{yyy}\eta^{33}
  -2\eta^{12}\eta_{23}+\eta^{13}\eta_{33})f_{xxy}-f_{xyy}^2\eta^{33}\notag
  \\ \label{eq:1} +(\eta^{12}\eta_{22}-2\eta^{13}\eta_{23})f_{xyy}
  +\eta_{22}\eta^{13}f_{yyy}+\eta^{11}(\eta_{22}\eta_{33}-\eta_{23}^2)=0.
\end{gather*}
In order to proceed, we need to solve the WDVV equation with respect to one
dependent variable. More generally, we need a criterion to choose
\emph{parametric} derivatives and \emph{principal} derivatives in WDVV
equations. Parametric derivatives are free variables, and principal derivatives
are functions of the parametric derivatives through the WDVV equations (the
terminology is standard, see \cite{M09,Seiler:FTDEqApCAl}). This is
accomplished by choosing a distinguished independent variable; without loss of
generality, we choose $t^2=x$.

Then, we choose as parametric derivatives third-order derivatives of the form
$f_{xim}$; the only remaining derivative $f_{yyy}$ is the principal
derivative.  It is easy to realize that, thanks to the nondegeneracy of~$\eta$,
the above WDVV equation can always be solved for $f_{yyy}$ provided we exclude
a singular set.

It is important to observe that the equations $S_{1223}=0$ and $S_{1233}=0$,
despite being identically zero, can be rewritten as $v^1 = u^2$ and
$v^2 = u^3$, respectively.

Now, let us consider first-order systems~\eqref{eq:8}: in this case
there is only one, namely
\begin{equation*}
  (u^1)_y = (v^1)_x,\quad (u^2)_y = (v^2)_x,\quad (u^3)_y = (v^3)_x.
\end{equation*}
The restriction of this system to the first prolongation
$\mathcal{E}^{(1)}\subset J_4E \subset \hat{J}_4E \subset J_1J_3E$ of the WDVV
equations $\mathcal{E}\subset J_3E$ is the quasilinear first-order system in
the unknown functions $u^1(x,y)$, $u^2(x,y)$, $u^3(x,y)$:
\begin{gather}\label{eq:52}
 (u^1)_y = (u^2)_x,
    \quad
     (u^2)_y = (u^3)_x,
    \quad
     (u^3)_y = \left(\frac{N}{\eta_{22}\eta^{13} + \eta^{23}u^1 + \eta^{33}u^2}\right)_x,
\end{gather}
where we used the WDVV equations~$S_{1223}=0$, $S_{1233}=0$ and $S_{2233}=0$
in order to write $v^1$, $v^2$ and $v^3$ as functions of $u^1$, $u^2$, $u^3$.
Here
\begin{gather*}
  N= - \eta_{22}\eta_{33}\eta^{11} + \eta_{23}^2\eta^{11}
  + (2\eta_{23}\eta^{12} - \eta_{33}\eta^{13})u^2
  + (2\eta_{23}\eta^{13} - \eta_{22}\eta^{12})u^3
     - \eta_{33}\eta^{12}u^1
    \\
     + \eta^{22}(u^2)^2 +
    \eta^{23}u^2u^3 + \eta^{33}(u^3)^2
    - \eta^{22}u^1u^3.
  \end{gather*}
Note that the denominator does not vanish on an open dense subset of the space
of dependent variables, in view of the nondegeneracy
of~$\eta$.

\begin{proposition}\label{pro:sol-wdvv}
  From any solution $(u^1,u^2,u^3)$ of system~\eqref{eq:52} it is possible to
  obtain a solution of the WDVV equation up to second degree
  polynomials.
\end{proposition}
\begin{proof}
  Indeed, one can introduce a potential function $h^1$ for the
  pair $(u^1,u^2)$ due to the first equation. Similarly, one can introduce
  potential functions $h^2$ of $(u^2,u^3)$ and $h^3$ of $(u^3,v^3)$.
  Then we have the equalities $h^1_y=h^2_x$ and $h^2_y=h^3_x$ by which we can
  introduce two further potentials, $g^1$ of $(h^1,h^2)$ and $g^2$ of
  $(h^2,h^3)$ and finally a potential $f$ of $(g^1,g^2)$.
\end{proof}
We note that the solution of WDVV equations that can be obtained by the above
procedure belongs to the non-singular part of the WDVV equations which is
selected by the condition that the denominator in~\eqref{eq:52} does not
vanish.

The calculation of this Subsection are available in the repository
\cite{opavit26:_maple_reduc_wdvv}.

\subsection{Main example: WDVV equations,
  \texorpdfstring{$\mathbf{N=4}$}{N=4}}
\label{sec:main-example-sys}

This case is much less studied. The first work is \cite{ferapontov96:_hamil}
(see also \cite{mokhov98:_sympl_poiss}), and it is carried out in the case
$\eta=\eta^{(1)}$. Here, we work with $\eta=\eta^{(2)}$ as it is unpractical to
present the calculations for a generic $\eta$; our calculations are valid also
in the case $\eta=\eta^{(1)}$ if $\mu=0$. We take the original viewpoint of
understanding the commutativity of the first-order WDVV systems as a
manifestation of the compatibility of the WDVV equations themselves.

Let us introduce the notation $(x,y,z)=(t^2,t^3,t^4)$. Following
Subsection~\ref{sec:wdvv-systems}, we introduce the new variables~\eqref{eq:50}
for third-order derivatives. We want to restrict equations \eqref{eq:6} to the
first prolongation of the WDVV equations. The equations \eqref{eq:6} read
\begin{equation}\label{eq:41}
  (u^I_x)_y = (u^I_y)_x,\quad (u^I_x)_z = (u^I_z)_x ,\quad (u^I_y)_z = (u^I_z)_y,
\end{equation}
where $I\in\{1,\ldots,6\}$ ($n=6$).
The WDVV equations are
\begin{equation}
  \label{eq:36}
\begin{split}
&\mu f_{yyz}f_{zzz}-\mu f_{yzz}^2 + 2f_{yyz}f_{xyz} -f_{yyy}f_{xzz} -f_{xyy}f_{yzz}  = 0,\\
&f_{xxy}f_{yzz} -f_{xxz}f_{yyz} -\mu f_{zzz}f_{xyz} + f_{zzz} + f_{xyy}f_{xzz} + \mu f_{xzz}f_{yzz} -f_{xyz}^2 = 0,\\
&f_{xxy}f_{yyz} -f_{xxz}f_{yyy} + \mu f_{yyz}f_{xzz} -\mu f_{xyz}f_{yzz} + f_{yzz} = 0,\\
&f_{xxy}f_{xzz} -\mu f_{xxz}f_{zzz} -2f_{xxz}f_{xyz} + f_{xxx}f_{yzz} + \mu f_{xzz}^2 = 0,\\
&f_{xxz}f_{xyy} + \mu f_{xxz}f_{yzz} -f_{yyz}f_{xxx} -\mu f_{xzz}f_{xyz} + f_{xzz} = 0,\\
&f_{xxy}f_{xyy} + \mu f_{xxz}f_{yyz} -f_{xxx}f_{yyy} -\mu f_{xyz}^2 + 2f_{xyz} = 0.
\end{split}
\end{equation}

Again, the choice of principal and parametric derivatives in the WDVV equations
depends on the choice of a distinguished independent variable; and again,
without loss of generality, we choose $t^2=x$. Then, parametric derivatives
have the form $f_{xim}$ and there are four principal derivatives: $f_{yyy}$,
$f_{yyz}$, $f_{yzz}$, $f_{zzz}$.

\begin{proposition}\label{prop:orthoN4}
  The WDVV equations~\eqref{eq:36} can be solved with respect to the principal
  derivatives $f_{yyy}$, $f_{yyz}$, $f_{yzz}$, $f_{zzz}$ on a dense open subset
  of the space of third-order derivatives of $f$. More precisely, the WDVV
  equations~\eqref{eq:36} are locally equivalent to a system of four nonlinear
  PDEs, each expressing one of the principal derivatives as a rational function
  of $f_{xxx}$, $f_{xxy}$, $f_{xyy}$, $f_{xxz}$, $f_{xyz}$, $f_{xzz}$.
\end{proposition}
\begin{proof}
  Indeed, it is possible to select a subsystem of the WDVV
  equations~\eqref{eq:36} which is linear with respect to the chosen principal
  derivatives (we recall that they are $x$-free derivatives); for example, we
  can choose the last $4$ equations.

  Following the numbering convention \eqref{eq:50} we introduce the letters
  $u^1=f_{xxx}$, $u^2=f_{xxy}$, $u^3=f_{xyy}$, $u^4=f_{xxz}$, $u^5=f_{xyz}$,
  $u^6=f_{xzz}$, and denote
\begin{equation*}
  S:= \mu^2 u^4(u^1 u^6 - (u^4)^2) +\mu u^1(u^2 u^4 - u^1 u^5) + (u^1)^2.
\end{equation*}
Then the solution of the above mentioned subsystem of~\eqref{eq:36} is
equivalent to the system
\begin{subequations}\label{eq:39}
\begin{align}
  \begin{split}
  & f_{yyy}=\frac{1}{S}\big( \mu^2 \left(u^2 u^3 u^4 u^6 - u^2 u^4 (u^5)^2 - u^3(u^4)^2 u^5 + u^1  (u^5)^3\right) + u^1 (u^2 u^3+ 2 u^5)
\\ &\hphantom{ciaociao}
  +\mu\left((u^2)^2 u^3 u^4+ 2 u^2 u^4 u^5 - u^1 u^2 u^3 u^5
 + u^3(u^4)^2 + u^4 u^6  - 3 u^1 (u^5)^2\right)\big),
\end{split}
  \\
\begin{split}
  & f_{yyz}=\frac{1}{S}\big( \mu^2\left(u^1 u^6 - (u^4)^2\right) (u^5)^2 + u^1(u^4  u^3+ u^6)\\
  &\hphantom{ciaociao}
  + \mu\left(u^2 u^3(u^4)^2+ 2 (u^4)^2 u^5 - u^1 u^3 u^4 u^5 - 2 u^1 u^5 u^6\right)\big),
\end{split}
  \\
  \begin{split}
  & f_{yzz}=\frac{1}{S}(
   \mu^2(u^1 u^6 -  (u^4)^2) u^5 u^6+ (2 u^4 u^5- u^2 u^6) u^1
\\
  &\hphantom{ciaociao}  +\mu\left( u^1 u^2 u^5 u^6  + u^3(u^4)^3 + (u^4)^2 u^6  - u^1 u^3 u^4 u^6
  - u^1 u^4 (u^5)^2 - u^1 (u^6)^2\right)
),
  \end{split}
  \\
  \begin{split}
    &f_{zzz}=\frac{1}{S}(\mu^2( u^1 u^6 - (u^4)^2) (u^6)^2
 +\mu\left( 2 u^1 u^2 (u^6)^2  - u^2 (u^4)^2 u^6 + 2 (u^4)^3 u^5- 3 u^1 u^4 u^5 u^6 \right)
\\
  & \hphantom{ciaociao}  - 2 u^1 u^2 u^4 u^5 + u^1 u^3(u^4)^2 + u^1 u^4 u^6 - (u^1)^2 u^3 u^6 + (u^1)^2 (u^5)^2+ u^1(u^2)^2 u^6).
  \end{split}
\end{align}
\end{subequations}
Then, it is easy to prove, for example by a computer algebra system, that the
first two equations of~\eqref{eq:36} identically vanish on account
of~\eqref{eq:39}, provided one excludes the singular variety $S=0$.
\end{proof}

We can give an original and more `structural' interpretation of the above
result.  We recall~\cite{M09,Seiler:FTDEqApCAl} that a system is in
\emph{orthonomic form} if every equation is solved with respect to one of the
highest (with respect to the chosen ordering) derivatives, and no such
derivatives appear in the right-hand side of any solved equation.
\begin{corollary}\label{coro:orthoN4}
  The WDVV system~\eqref{eq:36} can be locally written in orthonomic form.
\end{corollary}

The above viewpoint leads to an easy answer in the affirmative to the question:
\emph{is the WDVV system~\eqref{eq:36} compatible?} Such a question was termed
`a nontrivial exercise' in \cite{D96} (for the case $N=4$ and
$\eta=\eta^{(1)}$). `Compatibility' for orthonomic systems translates into
`passivity': the absence of hidden integrability conditions \cite{M09}.
\begin{proposition}\label{pro:passive}
  The local orthonomic form~\eqref{eq:39} of the WDVV system is passive: there
  are no hidden integrability conditions.
\end{proposition}
\begin{proof}
  By means of computer algebra it is easy to prove that the conditions
  \begin{equation*}
    (f_{yyy})_z = (f_{yyz})_y,\quad (f_{yyz})_z = (f_{yzz})_y,\quad
    (f_{yzz})_z = (f_{zzz})_y
  \end{equation*}
  are identically satisfied once we replace the derivative coordinates by their
  expressions in~\eqref{eq:39}. These are the only conditions that is necessary
  to check in order to guarantee that the WDVV system, in the
  form~\eqref{eq:39}, is passive, according with the algorithm in \cite{M09}.
\end{proof}

The above results have important consequences.
\begin{proposition}\label{pro:commuting}
  Let us denote $u^I:=u^I_x$, $I=1$ \dots, $6$. Then, the restriction of the
  equations~\eqref{eq:41} to the prolonged WDVV equations (in orthonomic
  form~\eqref{eq:36})
  \begin{equation}\label{eq:60}
    \mathcal{E}^{(1)}\subset J_4E \subset \hat{J}_4E \subset J_1J_3E
  \end{equation}
  is equivalent, on a dense open subset of the space of dependent variables, to
  two \emph{commuting} first-order systems of conservation laws of the form
    \begin{equation}
      \label{eq:42}
      (u^I)_y = (v^I)_x,\qquad (u^I)_z = (w^I)_x,\qquad I=1,\ldots,6
    \end{equation}
    where $v^I=v^I(u^k)$, $w^I=w^I(u^k)$, $k=1$, \dots, $6$.
    The further conditions are equivalent to $v^I_z=w^I_y$, and are identically
    verified.
\end{proposition}
\begin{proof}
  From the ranking \eqref{eq:50} we can introduce the variables
  $v^I=u^I_y$ and $w^I = u^I_z$, where
  \begin{gather*}
    v^1=f_{yxx},\quad v^2=f_{yxy},\quad v^3=f_{yyy},\quad
    v^4=f_{yxz},\quad v^5=f_{yyz},\quad v^6=f_{yzz},\\
    w^1=f_{zxx},\quad w^2=f_{zxy},\quad w^3=f_{zyy},\quad
    w^4=f_{zxz},\quad w^5=f_{zyz},\quad w^6=f_{zzz}.
  \end{gather*}
  Then, the equation $(u^I_y)_z = (u^I_z)_y$ in~\eqref{eq:41} is just
  $(v^I)_z=(w^I)_y$.  By inspection of the right-hand side we note that
  there are three trivial identities: $v^1_z=w^1_y$, $v^4_z=w^4_y$,
  $v^5_z=w^5_y$, while the restriction of the first prolongation of the WDVV
  equations to the other three relations are equivalent to the compatibility
  (or passivity) of the WDVV system.

  In each of the relations $(u^I)_y = (v^I)_x$ and $(u^I)_z = (w^I)_x$ there
  are $3$ trivial equations, in accordance with
  Proposition~\ref{pro:trivialeq}. In the remaining equations, $v^I$ and $w^I$
  can be expressed via the parametric derivatives, which we are denoting by
  $u^I$, $I=1$, \dots, $6$. This means that the restriction of~\eqref{eq:41} to
  the first prolongation of the WDVV equations yields two quasilinear
  first-order systems of conservation laws in the unknown functions $u^I(x,y,z)$
  and $u^I(x,y,z)$, respectively. In the system $(u^I)_y = (v^I)_x$ the
  variable $z$ plays the role of a parameter, and in the system $(u^I)_z =
  (w^I)_x$ the variable $y$ plays the role of a parameter.

  Commutativity is expressed, in geometric terms, by the fact that Jacobi
  bracket $[v,w]$ \cite{KrasilshchikVinogradov:SCLDEqMP,Seiler:FTDEqApCAl} of
  the two generalized vector fields $v=v^I_x\p_{u^I}$ and $w=w^I_x\p_{u^I}$ is
  zero. The expression of the bracket reads as
  \begin{equation*}
    [v,w]^I =  \mathrm D_\sigma v^J_x\pd{}{u^J_\sigma}(\mathrm D_xw^I) -
    \mathrm D_\sigma w^J_x\pd{}{u^J_\sigma}(\mathrm D_xv^I).
  \end{equation*}
  Using the systems of conservation laws~\eqref{eq:42} we can rewrite the
  bracket as
  \begin{align*}
    \notag
    [v,w]^i = &
    \left(\pd{}{y} + u^J_{\sigma y}\pd{}{u^J_\sigma}\right)\mathrm D_xw^I_x
    - \left(\pd{}{z} + u^J_{\sigma z}\pd{}{u^J_\sigma}\right)\mathrm D_xv^I_x
    \\ \notag
    = & \mathrm D_y w^I_x - \mathrm D_zv^I_x
    \\
    = & (w^I)_{xy} - (v^I)_{xz} = 0.
  \end{align*}
  The last equality holds due to the identity $(v^I)_z=(w^I)_y$ that has been
  proved above.
\end{proof}
Note that the above computation is available in the
repository~\cite{opavit26:_maple_reduc_wdvv}.

It is possible to generalize the procedure of Proposition~\ref{pro:sol-wdvv} to
this case.

\begin{proposition}\label{pro:equiv-wdvv-4}
  From any joint solution $(u^1,u^2,u^3,u^4,u^5,u^6)$ of the two
  systems~\eqref{eq:42} it is possible to obtain a solution of the WDVV
  equations up to second degree polynomials.
\end{proposition}
\begin{proof}
  Let us complete the solution with the further functions
  $u^7=f_{yyy}$, $u^8=f_{yyz}$, $u^9=f_{yzz}$, $u^{10}=f_{zzz}$, where the
  expression for the right-hand sides is given by the right-hand sides of
  equations~\eqref{eq:39}.

  It is clear that we can associate to every $u^I$ a multiindex of length $3$,
  say $\tau_3$. Accordingly, let us relabel the variables $u^I$ as
  $u^{\tau_3}$. Now, for every multiindex of length~$2$, say $\tau_2$, we
  define (up to a constant) an irrotational vector field
  $u^{\tau_2+\lambda}\pd{}{t^{\lambda}}$; correspondingly, we introduce the
  potential function $g^{\tau_2}$ such that its
  gradient is $g^{\tau_2}_{,\lambda}=u^{\tau_2+\lambda}$. The functions
  $g^{\tau_2}$ are defined up to a constant.

  Again, for every multiindex $\tau_1$ of length $1$, we can define the
  irrotational vector field $g^{\tau_1+\lambda}\pd{}{t^\lambda}$. Indeed,
  $g^{\tau_1+\lambda}_\mu = u^{\tau_1+\lambda+\mu} = g^{\tau_1+\mu}_\lambda$.
  Correspondingly, we introduce the potential function $h^{\tau_1}$.

  Finally, we introduce a function $f$ such that $f_{,\lambda}=h^{\tau_1}$,
  where $\tau_1=(\lambda)$. Indeed, the vector field
  $h^{(\lambda)}\pd{}{t^\lambda}$ is irrotational:
  $h^{(\lambda)}_{,\mu} = g^{(\lambda)+\mu} = h^{(\mu)}_{,\lambda}$.

  The function $f$ is determined up to second degree polynomials of the
  independent variables, as the values of the third-order derivatives are all
  given.
\end{proof}

Hence, the joint solutions of the two commuting systems are in correspondence
with the solutions of WDVV equations, as in the case $N=3$ (see
Proposition~\ref{pro:sol-wdvv}).

One of our aims is to generalize the above viewpoint and results as much as
possible.

For the sake of completeness, we observe that two quasilinear first-order
systems commute if and only if each of them is a generalized symmetry of the
other, or that each of them is in the kernel of the linearization of the other
\cite{KrasilshchikVinogradov:SCLDEqMP,Olver:ApLGDEq}.

\section{WDVV equations as Hamiltonian systems}
\label{sec:wdvv-equations-as}

In this Section we would like to prove the Structure Theorem, which is the
`heart' of the paper. Indeed, WDVV equations can be interpreted as families of
linear systems of equations in Pl\"ucker coordinates, with respect to the
Pl\"ucker embedding of a suitable projective space, discussed in
Subsection~\ref{sec:third-order-operator}.

Each linear systems of equations as above is of the type (studied
in~\cite{FPV17:_system_cl}) that is associated with systems of conservation
laws~\eqref{eq:3} with a Hamiltonian structure of the third
order~\eqref{DPHamOp}.  Indeed, by means of the Metric Theorem we will prove
that such Hamiltonian structures exist and are unique.

\subsection{Symmetries of WDVV equations and trivial cases}
\label{sec:symm-wdvv-equat}

We begin our analysis by an enumeration of the most
evident symmetry properties of the equations. This leads us to preliminary
results on the identification of trivial WDVV equations.

\begin{lemma}\label{lemma:syms}
  The WDVV equations fulfill the following symmetry properties:
  \begin{subequations}\label{eq:81}
  \begin{align}
    &S_{\alpha\beta\gamma\nu}=S_{\gamma\nu\alpha\beta},\label{eq:70}
    \\
    &S_{\alpha\beta\gamma\nu}=S_{\beta\alpha\nu\gamma},\label{eq:72}
    \\
    &S_{\alpha\beta\gamma\nu}= - S_{\alpha\nu\gamma\beta},\label{eq:73}
    \\ \label{eq:733}
    &S_{\alpha\beta\gamma\nu}=S_{\alpha\beta\nu\gamma} + S_{\alpha\gamma\beta\nu},
  \end{align}
  where all indices run from $1$ to $N$.
  \end{subequations}
\end{lemma}
We remind that the above symmetry properties are inherited by the fact that the
WDVV conditions can be interpreted as a zero curvature condition
(see~\cite{Mok99}).  In particular, the condition~\eqref{eq:733} is equivalent
to the Bianchi identity.

\begin{lemma}
  For any $\alpha$, $\beta$ and~$\gamma$ it holds that
  $S_{\alpha\beta\alpha\gamma}=0$ and $S_{\alpha\beta\gamma\beta}=0$
  identically. It turns out that any set of indices with at least~$3$ repeated
  indices yields an identically vanishing WDVV equation.
\end{lemma}

\begin{lemma}
  Any set of indices with exactly two repeated indices yields only one WDVV
  equation (up to an overall sign).
\end{lemma}
\begin{proof}
  There are twelve distinct sets of indices constructed from the set
  $\{\alpha,\alpha,\beta,\gamma\}$; using~\eqref{eq:81} we easily realize that
  the corresponding equations are the same up to a sign.
\end{proof}

So, nontrivial equations are of one of the following types:
\begin{enumerate}
\item $S_{\alpha\alpha\beta\gamma}=0$, where $\alpha$, $\beta$, $\gamma$ are
  distinct and $\beta\leq\gamma$, and
\item $S_{\alpha\beta\gamma\nu}=0$, where all indices are distinct.
\end{enumerate}
\begin{lemma}
  The subsystem of WDVV equations obtained by means of a set of mutually
  distinct indices $\{\alpha,\beta,\gamma,\nu\}$ is equivalent to two
  independent equations
  \begin{equation*}
    S_{\alpha\beta\gamma\nu} = 0\quad\text{and}\quad
    S_{\alpha\beta\nu\gamma} = 0,
  \end{equation*}
  where $\alpha<\beta<\gamma<\nu$.
\end{lemma}
\begin{proof}
It is easy to realize that we can always move a specified index in
$S_{\alpha\beta\gamma\nu}$ to the first place (up to a sign),
\begin{equation*}
  S_{\alpha\beta\gamma\nu} = S_{\beta\alpha\nu\gamma} =
  S_{\gamma\nu\alpha\beta} = - S_{\nu\alpha\beta\gamma}.
\end{equation*}
Moving to the beginning the smaller index $\alpha$, we remain with $6$ cases:
\begin{gather*}
  S_{\alpha\beta\gamma\nu} = 0,\quad S_{\alpha\gamma\beta\nu}=0,\quad
  S_{\alpha\nu\beta\gamma} = 0,
  \\
  S_{\alpha\beta\nu\gamma}=0,\quad
    S_{\alpha\nu\gamma\beta} = 0,\quad S_{\alpha\gamma\nu\beta}=0,
  \end{gather*}
  which can be further reduced to
  \begin{equation*}
    S_{\alpha\beta\gamma\nu} = 0,\quad S_{\alpha\beta\nu\gamma}=0,
    \quad S_{\alpha\gamma\beta\nu}=0.
  \end{equation*}
  using~\eqref{eq:73}. In view of~\eqref{eq:733} we have the result.
\end{proof}

In order to further analyze the WDVV equations we
introduce Latin indices $i$, $j$, $k$,\dots running
from~$2$ to~$N$. We have the expressions
\begin{equation}\label{eq:51}
\begin{split}
  &F_1=\frac{1}{2}\eta_{11}(t^1)^2 +
  \eta_{1k}t^kt^1 + \frac{1}{2} \eta_{sk}t^st^k,
  \\
  &F_i=\frac{1}{2}\eta_{1i}(t^1)^2 + \eta_{i k}t^kt^1 + f_i,
\\
&F_{11}=\eta_{11}t^1 + \eta_{1k}t^k,
\\
&F_{1i}=\eta_{1i}t^1 + \eta_{i k}t^k.
\end{split}
\end{equation}

\begin{lemma}
  The expression $S_{11ij}$ vanishes identically.
\end{lemma}
\begin{proof}
  \begin{align*}
    S_{11ij} =& \eta^{\lambda\mu}(F_{\lambda 11}F_{\mu ij} -
                F_{\lambda 1 j}F_{\mu i1})
    \\
    =& \eta^{11}(F_{111}F_{1ij} -
       F_{11 j}F_{1i1}) +
       \eta^{1k}(F_{1 11}F_{k ij} -
       F_{1 1 j}F_{k i1})
       \\
     & + \eta^{k 1}(F_{k 11}F_{1 ij} -
       F_{k 1 j}F_{1 i1})
     + \eta^{hk}(F_{h 11}F_{k ij} -
       F_{h 1 j}F_{k i1})
    \\
    =& \eta^{11}(\eta_{11}\eta_{ij} -
       \eta_{1j}\eta_{1i}) +
       \eta^{1k}(\eta_{11}f_{kij} -
       \eta_{1 j}\eta_{k i})
       \\
     & + \eta^{k1}(\eta_{1k}\eta_{ij} -
       \eta_{kj}\eta_{1i})
     + \eta^{hk}(\eta_{1h}f_{kij} -
       \eta_{hj}\eta_{ki})
    \\
    =&(\eta^{11}\eta_{11} + \eta^{k1}\eta_{1k})\eta_{ij} - (\eta^{11}\eta_{1j}
       + \eta^{k1}\eta_{kj})\eta_{1i}
    \\
              & + (\eta^{1k}\eta_{11} +\eta^{hk}\eta_{1h}) f_{kij}
                - (\eta^{1k}\eta_{1j} + \eta^{hk}\eta_{hj})\eta_{ki}
    \\
    =& \eta_{ij} - \delta^k_j\eta_{ki} = 0.
  \end{align*}
\end{proof}

\begin{lemma}\label{lem:triveq}
  The expression $S_{1kij}$ vanishes identically.
\end{lemma}
\begin{proof}
  \begin{align*}
    S_{1kij} =& \eta^{\lambda\mu}(F_{\lambda 1k}F_{\mu ij} -
                F_{\lambda 1 j}F_{\mu ik})
    \\
    =& \eta^{11}(F_{11k}F_{1ij} -
       F_{11 j}F_{1ik}) +
       \eta^{1k}(F_{1 1k}F_{k ij} -
       F_{1 1 j}F_{k ik})
       \\
     & + \eta^{k 1}(F_{k 1k}F_{1 ij} -
       F_{k 1 j}F_{1 ik})
     + \eta^{hk}(F_{h 1k}F_{k ij} -
       F_{h 1 j}F_{k ik})
    \\
    =& \eta^{11}(\eta_{1k}\eta_{ij} -
       \eta_{1j}\eta_{ik}) +
       \eta^{1k}(\eta_{1k}f_{kij} -
       \eta_{1 j}f_{k ik})
       \\
     & + \eta^{k1}(\eta_{kk}\eta_{ij} -
       \eta_{kj}\eta_{ik})
     + \eta^{hk}(\eta_{hk}f_{kij} -
       \eta_{hj}f_{kik})
    \\
    =&(\eta^{11}\eta_{1k} + \eta^{k1}\eta_{kk})\eta_{ij} - (\eta^{11}\eta_{1j}
       + \eta^{k1}\eta_{kj})\eta_{ik}
    \\
              & + (\eta^{1k}\eta_{1k} +\eta^{hk}\eta_{hk}) f_{kij}
                - (\eta^{1k}\eta_{1j} + \eta^{hk}\eta_{hj})f_{kik}
    \\
    =& f_{kij} - f_{jik} = 0,
  \end{align*}
where the symmetry of third partial derivatives is used in the last line.
\end{proof}
Since if there is~1 in the index of an equation we can always move it to the
first spot by means of symmetries, in view of the above results we can take
the remaining indices to be running from~2 to~$N$.

With the help of symmetries given in Lemma~\ref{lemma:syms} we can reduce
any equation with a repeating index to a canonical form~$S_{iijk}$ with
$j\leqslant k$, and any equation with no repeating indices to one of two
canonical forms~$S_{ijkl}$ or $S_{ijlk}$ with $i<j<k<l$.

We are now ready to state the Structure Theorem. We introduce the statement by
means of the usual low-dimensional examples.

\subsection{A particular case: WDVV equations,
  \texorpdfstring{$\mathbf{N=3}$}{N=3}}
\label{sec:part-case:-wdvv}

The first result showing that first-order WDVV systems possess a third-order
Hamiltonian formulation dates back to \cite{FGMN97}
($\eta=\eta^{(1)}$). Further results were obtained in
\cite{kalayci98:_alter_hamil_wdvv,kalayci97:_bi_hamil_wdvv}. In
\cite{vasicek21:_wdvv_hamil} it was established, using $\eta^{(1)}$ and
$\eta^{(2)}$ and invariance, that such a structure exists in general, when
$N=3$. Here, we give a direct proof of this fact with explicit expression for
the Hamiltonian structure for every choice of $\eta$.

When $N=3$ there is only one equation which is nontrivial, namely $S_{2233}=0$.
We can add two further trivial equations, $S_{1223}=0$ and $S_{1233}=0$, and
index the three equations with $\gamma=3$, $1$, $2$ in the above order. We
recall that we have the system of conservation laws
$(u^I)_y = (v^I)_x$, $I=1$, $2$, $3$, cf.~\eqref{eq:52}.

Then, the three WDVV equations $S_{2233}=0$, $S_{1223}=0$ and $S_{1233}=0$ have
the form of a linear line congruence in the Pl\"ucker's coordinates described
in Subsection~\ref{sec:third-order-operator}, see~\eqref{eq:9},
\begin{equation}\label{eq:22}
\psi^\gamma_{km}(u^mv^k-u^kv^m)+\omega^\gamma_kv^k-\theta^\gamma_mu^m-\xi^\gamma=0,
\end{equation}
where (in the expression $\psi^\gamma_k$ and alike, $\gamma$ is the row, $k$ is
the column)
\begin{gather*}
(\psi^1_{kh})=(\psi^2_{kh})=0,
\quad
(\psi^3_{kh})=
\begin{pmatrix} 0 & -\eta^{22} & -\eta^{23}\\
\eta^{22} &0&-\eta^{33}\\ \eta^{23}&\eta^{33}&0
\end{pmatrix},  \\[1ex]
\Omega=(\omega^\gamma_k)=
\begin{pmatrix}
1 &0&0\\
0&1&0\\
-\eta^{12}\eta_{23} & \eta^{12}\eta_{22}-\eta^{13}\eta_{23} & \eta^{13}\eta_{22}
\end{pmatrix},
\\[1ex]
\Theta=(\theta^\gamma_k)=
\begin{pmatrix}
0 &1&0\\
0&0&1\\
-\eta^{12}\eta_{33} & \eta^{12}\eta_{23}-\eta^{13}\eta_{33} & \eta^{13}\eta_{23}
\end{pmatrix},\quad
\xi = (\xi^\gamma) =
\begin{pmatrix}
 0 \\ 0 \\ -\eta^{11}(\eta_{22}\eta_{33} - (\eta_{23})^2)
\end{pmatrix}
\\[1ex]
\Phi=(\phi_{\alpha\beta})=
\begin{pmatrix}
\eta^{11}\eta^{22}-(\eta^{12})^2&\eta^{11}\eta^{23}-\eta^{12}\eta^{13}&\eta^{12}\eta^{23}-\eta^{13}\eta^{22}\\
\eta^{11}\eta^{23}-\eta^{12}\eta^{13}&\eta^{11}\eta^{33}-(\eta^{13})^2&\eta^{12}\eta^{33}-\eta^{13}\eta^{23}\\
\eta^{12}\eta^{23}-\eta^{13}\eta^{22} & \eta^{12}\eta^{33}-\eta^{13}\eta^{23}&\eta^{22}\eta^{33}-(\eta^{23})^2
\end{pmatrix}.
\end{gather*}

We have the following results.
\begin{theorem}
  Let $N=3$, and choose one of the independent variables $t^2$ or $t^3$. Then:
  \begin{itemize}
  \item The matrix $\phi$ is nondegenerate.
  \item The matrix $(\psi^\gamma_i)$, where $\psi^\gamma_i =
    \psi^\gamma_{ij}u^j + \omega^\gamma_k$, is nondegenerate on an open dense
    subset of the space of dependent variables.
  \item The above equations of a linear line congruence can be rewritten as a
    system of conservation laws, which is Hamiltonian with respect to a
    third-order Hamiltonian operator.
  \end{itemize}
\end{theorem}
\begin{proof}
  Let us choose $t^2$. It is easy to check the first two statements by a
  computer algebra system. In particular, $\phi$ turns out to be the metric
  which is naturally induced on $\wedge^2\mathbb{R}^3$ by $\eta$. Moreover, we
  have
  \begin{displaymath}
    \det(\psi^\gamma_i) = \eta_{22}\eta^{13} + \eta^{23}u^1 + \eta^{33}u^2,
  \end{displaymath}
  which is just the denominator in~\eqref{eq:52} and is not identically zero
  due to the nondegeneracy of $\eta$.

  Then, the constants $\phi_{\alpha\beta}$, $\psi^\gamma_i$,
  $\omega ^\gamma_k$ fulfill the system~\eqref{eq:48}, thus
  yielding a third-order Hamiltonian operator $A_2$ of the form \eqref{DPHamOp}
  defined by the Monge metric
  $f_{ij}=\phi_{\alpha\beta}\psi^\alpha_i \psi^\beta_j$ (see~\eqref{eq:30}).

  Moreover, recalling that the equations~\eqref{eq:9} are the expanded version
  of the equation $\psi^\gamma_i v^i = Z^\alpha$, where
  $Z^\alpha=\theta^\alpha_k u^k + \xi^\alpha$, we obtain the formula
  $v^i = \psi^i_\alpha Z^\alpha$. This, together with the fact that the
  constants $\phi_{\alpha\beta}$, $\psi^\gamma_{kh}$, $\omega^\gamma_k$,
  $\theta^\gamma_k$, $\xi^\gamma$ also fulfill the system~\eqref{eq:53}, proves
  that the system~\eqref{eq:52} is Hamiltonian with respect to the above
  third-order Hamiltonian operator $A_2$.

  If we choose $t^3$ as an independent variable, we observe that the above
  results continue to hold by exchanging the role of $u^I$ and $v^I$ in the
  above formula \eqref{eq:22}. Indeed, the independence from the choice of the
  variable $t^p$ is consistent with the formulae for the inversion
  $t^2 \leftrightarrow t^3$ in the system of conservation laws
  $(u^I)_{t^3}=(v^I)_{t^2}$ as stated in \cite{FPV17:_system_cl}.
\end{proof}
\begin{remark}\label{rem:Hamiltonian}
  For every Hamiltonian operator and quasilinear system of conservation laws as
  above we have an explicit formula for the Hamiltonian
  (see~\cite{FPV17:_system_cl}).
\end{remark}
\begin{remark}\label{rem:singular}
  The variety $\det(\psi^\gamma_i)=0$ is the \emph{singular surface} of the
  quadratic line complex corresponding to the Monge metric $f$
  \cite{FPV14,FPV16}.
\end{remark}

\subsection{Main example: WDVV equations,
  \texorpdfstring{$\mathbf{N=4}$}{N=4}}
\label{sec:main-example:-wdvv}

The Hamiltonian properties in dimension $N=4$ were much more difficult to
prove. The first results were in \cite{ferapontov96:_hamil}, but the
Hamiltonian structure that was constructed was a (local) first-order one. A
third-order Hamiltonian structure was constructed, in the case
$\eta=\eta^{(1)}$, in \cite{PV15}. This was extended to $\eta^{(2)}$ in
\cite{vasicek21:_wdvv_hamil} and to both canonical forms of $\eta$ in dimension
$N=5$. In this Subsection, we will show that these results are general in
dimension $N=4$: they hold for every choice of $\eta$ and for every choice of
the distinguished variable $t^p$. That will lead to our general results for
arbitrary $N$ in next Section.

According with the notation scheme of~\eqref{eq:50} (see also
Subsection~\ref{sec:main-example-sys}), we have three distinct group of
variables: $u^I_x$, $u^I_y$, $u^I_z$, $I=1$, \dots, $6$. Let us introduce the
notation $u^I=u^I_x$, $v^I=u^I_y$, $w^I=u^I_z$.

It is interesting to observe that the notation for $u^1$, $u^2$, $u^3$ and
$v^1$, $v^2$, $v^3$ is preserved from the case $N=3$, thus facilitating
proofs by induction. Of course, the letters $w^I$ represent new derivative
variables that are introduced when $N$ goes from $3$ to $4$. We have
\begin{gather*}
u^4=f_{xxz},\quad u^5=f_{xyz},\quad u^6=f_{xzz},\\
v^4=f_{xyz}=u^5,\quad v^5=f_{yyz},\quad v^6=f_{yzz},\\
w^1=f_{xxz}=u^4,\quad w^2=f_{xyz}=u^5,\quad w^3=f_{yyz},\quad
w^4=f_{xzz}=u^6,\quad w^5=f_{yzz},\quad w^6=f_{zzz}.
\end{gather*}

There are six non-trivial WDVV equations $S_{2233}=0$, $S_{2234}=0$,
$S_{2244}=0$, $S_{2334}=0$, $S_{2344}=0$, $S_{3344}=0$. We can supplement them
with suitable trivial equations in order to obtain three sets of six linear
relations in Pl\"ucker's coordinates with respect to three distinct Pl\"ucker's
embeddings~\eqref{eq:9}. These can be interpreted as three different linear
line congruences.

\textbf{First linear line congruence.} Let us label the equations $S_{2233}=0$,
$S_{2234}=0$, $S_{2334}=0$, supplemented with the trivial equations $S_{1223} =
u^2 - v^1 =0$, $S_{1233} =u^3 - v^2=0$, $S_{1234} = u^5-v^4 = 0$, by
the index $\gamma=3,5,6,1,2,4$, resp. Then, the previous system can be written as
\[
  \frac12\psi^\gamma_{km}(u^mv^k-u^kv^m)+{_x\omega}^\gamma_kv^k
  - {_y\omega}^\gamma_mu^m-{_{xy}\xi}^\gamma=0,\quad \gamma=1,\ldots,6,
\]
where, with the same notation as in the previous Subsection,
\begin{gather*}
\psi^1_{kh}=\psi^2_{kh}=0,\quad (\psi^3_{kh})=
\begin{pmatrix}
0 & -\eta^{22} & -\eta^{23}& 0 & -\eta^{24} & 0\\
\eta^{22} &0&-\eta^{33}& \eta^{24}&-\eta^{34}&0\\
\eta^{23} &\eta^{33}&0& \eta^{34}&0&0\\
0 &-\eta^{24}&-\eta^{34}&0&-\eta^{44}&0\\
\eta^{24} &\eta^{34} &0& \eta^{44}&0&0\\
0 &0& 0 &0&0&0
\end{pmatrix},
\quad
\psi^4_{kh}=0,
\\[1ex]
(\psi^5_{kh})=
\begin{pmatrix}
0 & 0 & 0& -\eta^{22} & -\eta^{23} & -\eta^{24}\\
0 &0& 0&-\eta^{23}&-\eta^{33}&-\eta^{34}\\
0 &0&0 &0&0&0\\
\eta^{22} &\eta^{23}&0 &0&-\eta^{34}&-\eta^{44}\\
\eta^{23} &\eta^{33} &0&\eta^{34}&0&0\\
\eta^{24} &\eta^{34} &0&\eta^{44}&0&0
\end{pmatrix},
\quad
(\psi^6_{kh})=
\begin{pmatrix}
0 & 0 & 0 & 0 & 0 & 0\\
0 &0&0& -\eta^{22}&-\eta^{23}&-\eta^{24}\\
0 &0&0& -\eta^{23}&-\eta^{33}&-\eta^{34}\\
0 &\eta^{22}&\eta^{23}&0&\eta^{24}&0\\
0 &\eta^{23} &\eta^{33}& -\eta^{24}&0&-\eta^{44}\\
0 &\eta^{24} &\eta^{34}& 0&\eta^{44}&0
\end{pmatrix},
\\[1ex]
({_x\omega}^\gamma_k)=
\begin{pmatrix}
1 &0& 0 &0&0&0\\
0 &1& 0 &0&0&0\\
-\eta^{12}\eta_{23} & \eta^{12}\eta_{22}-\eta^{13}\eta_{23} &\eta^{13}\eta_{22} & -\eta^{14}\eta_{23}& \eta^{14}\eta_{22} & 0\\
0 &0& 0 &1&0&0\\
-\eta^{12}\eta_{24}  & -\eta^{13}\eta_{24} &0 &\eta^{12}\eta_{22}-\eta^{14}\eta_{24}& \eta^{13}\eta_{22} &\eta^{14}\eta_{22}\\
0 &-\eta^{12}\eta_{24}&-\eta^{13}\eta_{24} &\eta^{12}\eta_{23} &\eta^{13}\eta_{23}-\eta^{14}\eta_{24} & \eta^{14}\eta_{23}
\end{pmatrix},
\\[1ex]
({_y\omega}^\gamma_k)=
\begin{pmatrix}
0 &1& 0 &0&0&0\\
0 &0& 1 &0&0&0\\
-\eta^{12}\eta_{33} & \eta^{12}\eta_{23}-\eta^{13}\eta_{33} & \eta^{13}\eta_{23} & -\eta^{14}\eta_{33} &\eta^{14}\eta_{23}& 0\\
0 &0& 0 &0&1&0\\
-\eta^{12}\eta_{34} &-\eta^{13}\eta_{34}& 0&\eta^{12}\eta_{23}-\eta^{14}\eta_{34}&\eta^{13}\eta_{23}&\eta^{14}\eta_{23}\\
0 &-\eta^{12}\eta_{34}&-\eta^{13}\eta_{34} &\eta^{12}\eta_{33}&\eta^{13}\eta_{33}-\eta^{14}\eta_{34}&\eta^{14}\eta_{33}
\end{pmatrix},
\\[1ex]
({_{xy}\xi}^\gamma)=
-\eta^{11}\begin{pmatrix}
0&
0&
\eta_{22}\eta_{33}-(\eta_{23})^2&
0&
\eta_{22}\eta_{34}-\eta_{23}\eta_{24}&
\eta_{23}\eta_{34}-\eta_{24}\eta_{33}
\end{pmatrix}^\intercal.
\end{gather*}

\textbf{Second linear line congruence.} The system $S_{1224}=u^4-w^1=0$, $S_{1234}=u^5-w^2=0$, $S_{1244}=u^6-w^4=0$,
$S_{2234}=0$, $S_{2244}=0$, $S_{2344}=0$ ($\gamma=1,2,4,3,5,6$, resp.)
can be written as
\[
  \frac12\psi^\gamma_{km}(u^mw^k-u^kw^m)+{_x\omega}^\gamma_kw^k
  -{_z\omega}^\gamma_mu^m-{_{xz}\xi}^\gamma=0,\quad \gamma=1,\ldots,6,
\]
where $\psi^\gamma_{kh}$ and ${_x\omega}^\gamma_k$ are the same as in the first embedding and
\begin{gather*}
({_z\omega}^\gamma_k)=
\begin{pmatrix}
0 &0& 0 &1&0&0\\
0 &0& 0 &0&1&0\\
-\eta^{12}\eta_{34} & \eta^{12}\eta_{24}-\eta^{13}\eta_{34} & \eta^{13}\eta_{24} & -\eta^{14}\eta_{34} &\eta^{14}\eta_{24}& 0\\
0 &0& 0 &0&0&1\\
-\eta^{12}\eta_{44} &-\eta^{13}\eta_{44}& 0&\eta^{12}\eta_{24}-\eta^{14}\eta_{44}&\eta^{13}\eta_{24}&\eta^{14}\eta_{24}\\
0 &-\eta^{12}\eta_{44}&-\eta^{13}\eta_{44} &\eta^{12}\eta_{34}&\eta^{13}\eta_{34}-\eta^{14}\eta_{44}&\eta^{14}\eta_{34}\\
\end{pmatrix}
\\[1ex]
({_{xz}\xi}^\gamma)=
-\eta^{11}\begin{pmatrix}
0&
0&
\eta_{22}\eta_{34}-\eta_{23}\eta_{24}&
0&
\eta_{22}\eta_{44}-(\eta_{24})^2&
\eta_{23}\eta_{44}-\eta_{24}\eta_{34}
\end{pmatrix}^\intercal.
\end{gather*}

\textbf{Third linear line congruence.} The system $S_{1324}=v^4-w^2=0$, $S_{1334}=v^5-w^3=0$, $S_{1344}=v^6-w^5=0$,
$S_{2334}=0$, $S_{2344}=0$, $S_{3344}=0$ ($\gamma=1,2,4,3,5,6$, resp.)
can be written as
\[
  \frac12\psi^\gamma_{km}(v^mw^k-v^kw^m)+{_y\omega}^\gamma_kw^k
  -{_z\omega}^\gamma_kv^k-{_{yz}\xi}^\gamma=0,\quad \gamma=1,\ldots,6,
\]
where $\psi^\gamma_{kh}$, ${_y\omega}^\gamma_k$, ${_z\omega}^\gamma_k$ are the same as before, and
\begin{gather*}
({_{yz}\xi}^\gamma)=
-\eta^{11}\begin{pmatrix}
0&
0&
\eta_{23}\eta_{34}-\eta_{24}\eta_{33}&
0&
\eta_{22}\eta_{44}-\eta_{24}\eta_{34}&
\eta_{33}\eta_{44}-(\eta_{34})^2
\end{pmatrix}^\intercal.
\end{gather*}

Let us introduce the matrix
\small
\begin{multline*}
\phi=(\phi_{\alpha\beta})=
\\
  \begin{pmatrix}
\eta^{11}\eta^{22}-(\eta^{12})^2 & \eta^{11}\eta^{23}-\eta^{12}\eta^{13} &
\eta^{12}\eta^{23}-\eta^{13}\eta^{22}& \eta^{11}\eta^{24}-\eta^{12}\eta^{14}& \eta^{12}\eta^{24}-\eta^{14}\eta^{22}& \eta^{13}\eta^{24}-\eta^{14}\eta^{23}
\\
\eta^{11}\eta^{23}-\eta^{12}\eta^{13}&\eta^{11}\eta^{33}-(\eta^{13})^2&
\eta^{12}\eta^{33}-\eta^{13}\eta^{23}& \eta^{11}\eta^{34}-\eta^{13}\eta^{14}& \eta^{12}\eta^{34}-\eta^{14}\eta^{23}& \eta^{13}\eta^{34}-\eta^{14}\eta^{33}
\\
\eta^{12}\eta^{23}-\eta^{13}\eta^{22}& \eta^{12}\eta^{33}-\eta^{13}\eta^{23}&
\eta^{22}\eta^{33}-(\eta^{23})^2 & \eta^{12}\eta^{34}-\eta^{13}\eta^{24}& \eta^{22}\eta^{34}-\eta^{23}\eta^{24}& \eta^{23}\eta^{34}-\eta^{24}\eta^{33}
\\
\eta^{11}\eta^{24}-\eta^{12}\eta^{14}& \eta^{11}\eta^{34}-\eta^{13}\eta^{14}&
\eta^{12}\eta^{34}-\eta^{13}\eta^{24} &\eta^{11}\eta^{44}-(\eta^{14})^2&\eta^{12}\eta^{44}-\eta^{14}\eta^{24} &\eta^{13}\eta^{44}-\eta^{14}\eta^{34}
\\
\eta^{12}\eta^{24}-\eta^{14}\eta^{22}& \eta^{12}\eta^{34}-\eta^{14}\eta^{23}&
\eta^{22}\eta^{34}-\eta^{23}\eta^{24}& \eta^{12}\eta^{44}-\eta^{14}\eta^{24} &\eta^{22}\eta^{44}-(\eta^{24})^2 &\eta^{23}\eta^{44}-\eta^{24}\eta^{34}
\\
\eta^{13}\eta^{24}-\eta^{14}\eta^{23}&\eta^{13}\eta^{34}-\eta^{14}\eta^{33}&
\eta^{23}\eta^{34}-\eta^{24}\eta^{33}& \eta^{13}\eta^{44}-\eta^{14}\eta^{34}& \eta^{23}\eta^{44}-\eta^{24}\eta^{34}&\eta^{33}\eta^{44}-(\eta^{34})^2
\end{pmatrix}
\end{multline*}
\normalsize

Then, the above linear line congruences have the following interpretation.
\begin{theorem}
  Let $N=4$, and choose an independent variable $t^p$, $p\geq 2$. Then,
  \begin{itemize}
  \item The matrix $\phi$ is nondegenerate.
  \item The matrix $(\psi^\gamma_i)$, where
    $\psi^\gamma_i=\psi^\gamma_{ik}u^k_p + {_p\omega^k_{i}}$, is
    nondegenerate, provided we exclude a dense open subset of the space of
    dependent variables.
  \item The two linear line congruences involving the coordinates
    $(u^I_p)$ can be rewritten in the form of two Hamiltonian systems of
    conservation laws:
    \begin{equation}
      \label{eq:58}
      (u^I_p)_{q_1} = (u^I_{q_1})_p,\qquad (u^I_p)_{q_2} = (u^I_{q_2})_p,
    \end{equation}
    where $\{q_1,q_2\} = \{2,3,4\}\setminus \{p\}$, where
    \begin{gather}\label{eq:59}
      u^I_{q_1} = \psi^I_\gamma Z^{\gamma}_{q_1},\qquad
      Z^{\gamma}_{q_1} = {}_{q_1}\omega^\gamma_k u^k_p + {}_{q_1p}\xi^\gamma,
      \\
      u^I_{q_2} = \psi^I_\gamma Z^{\gamma}_{q_2},\qquad
      Z^{\gamma}_{q_2} = {}_{q_2}\omega^\gamma_k u^k_p + {}_{q_2p}\xi^\gamma.
    \end{gather}
  \item The two systems \eqref{eq:58} are Hamiltonian with respect to
    the same third-order Hamiltonian operator in canonical form, defined by the
    Monge metric $f_{ij}=\phi_{\alpha\beta}\psi^\alpha_i\psi^\beta_j$.
  \item The two systems \eqref{eq:58} commute.
  \end{itemize}
\end{theorem}
\begin{proof}
  The proof is a generalization of the case $N=3$, with the difference that
  there will be $2$ systems of conservation laws. A direct inspection shows
  that the matrix $(\psi^\gamma_i)$ is the same for a given choice of~$t^p$,
  for both other independent variables; $\phi$ remains the same for all
  independent variables.

  We observe that, as in the case $N=3$, $\phi$ turns out to be the metric
  which is naturally induced on~$\wedge^2\mathbb{R}^4$ by $\eta$.

  Also in this case, $\det(\psi^\gamma_i)$ does not vanish identically; its
  zero locus is the singular surface of the quadratic line complex
  corresponding to the Monge metric $f$ \cite{FPV14,FPV16} (see
  Remark~\ref{rem:singular}).

  Commutativity is a slightly more general version of
  Proposition~\ref{pro:commuting}, and it again comes from the passivity of the
  WDVV system in orthonomic form, as Proposition~\ref{pro:passive}; passivity
  can be proved by means of computer algebra.

  As in the case $N=3$, if we choose another independent variable the
  above results continue to hold by exchanging the role of the corresponding
  set of dependent variables. Again, the independence from the choice of the
  variable $t^p$ is consistent with the formulae for the inversion
  $t^p \leftrightarrow t^q$ in the system of conservation laws
  $(u^I)_{t^q}=(v^I)_{t^p}$ as stated in \cite{FPV17:_system_cl}.
\end{proof}
As in the case $N=3$ (see Remark~\ref{rem:Hamiltonian}), for every Hamiltonian
operator and quasilinear system of conservation laws as above, we have an
explicit formula for the Hamiltonian (see~\cite{FPV17:_system_cl}).

\subsection{The Structure Theorem}

In this Subsection we will provide the Structure Theorem, \emph{i.e.}\,we will
give an inductive proof, with respect to $N$, of the fact that WDVV equations
have the form \eqref{eq:31}. The fact that such equations are indeed
Hamiltonian systems is dealt with in the next Section.

We consider a subsystem of the WDVV system that is made of nontrivial
equations
\begin{displaymath}
  S_{aabb}=0,\ S_{aabc}=0,\ S_{abbc}=0,\ S_{abcc}=0,\ S_{abcd}=0,\
  S_{abdc}=0,
\end{displaymath}
where $2\leqslant a<b<c<d\leqslant N$, having thus in total
\[\binom{N-1}{2}+3\binom{N-1}{3}+2\binom{N-1}{4}=\frac1{12}N(N-1)^2(N-2)\]
equations. We will join selected nontrivial equations to selected trivial
equations of the form of Lemma~\ref{lem:triveq}.

\begin{theorem}[Structure Theorem]\label{theor:WDVVLinComplexForm} Let
  $N\geq 3$.
  Given $2\leqslant \lambda<\mu\leqslant N$, the system
  \begin{gather*}
    \{S_{\lambda \beta \mu \nu},\ 1\leqslant \beta<\nu\leqslant N\}
  \end{gather*}
  can be written as
  \begin{gather}\label{eq:LinearComplexForm}
    \frac12\psi^\gamma_{km}(u_\lambda^mu_\mu^k-u_\lambda^ku_\mu^m)
    +{_\lambda\omega}^\gamma_ku_\mu^k-{_\mu\omega}^\gamma_mu_\lambda^m
    -{_{\lambda\mu}\xi}^\gamma=0,
  \end{gather}
  where the equation~$S_{\lambda \beta\mu\nu}$ is associated with $\gamma=(\nu,\beta+1)$
  and all coefficients are zeroes except:
  \begin{itemize}
  \item for $2\leqslant \beta<\nu\leqslant N$ and $2\leqslant a,b\leqslant N$
    we set
    \begin{subequations}\label{eq:14}
    \begin{equation}\label{eq:11}
        \psi^{(\nu,\beta+1)}_{(\nu,a)(\beta,b)}=\eta^{ab}
        = - \psi^{(\nu,\beta+1)}_{(\beta,b)(\nu,a)}
        \qquad\text{if $(\nu,a)\neq(\beta,b)$;}
    \end{equation}
  \item for $2\leqslant \beta<\nu< M\leq N$ and $\rho\in\{\lambda,\mu\}$ we set
    \begin{equation}\label{eq:18}
      {_\rho\omega}^{(\nu,\beta+1)}_{(M,\beta)}=-\eta^{M1}\eta_{\rho\nu},\quad
      {_\rho\omega}^{(\nu,\beta+1)}_{(M,\nu)}=\eta^{M1}\eta_{\rho\beta};
    \end{equation}
  \item for $2\leqslant \beta< \nu\leq N$ and $2\leq a<\nu$, with
    $\rho\in\{\lambda,\mu\}$, we set
    \begin{equation}\label{eq:24}
      {_\rho\omega}^{(\nu,\beta+1)}_{(\nu,a)}=
      \eta^{1a}\eta_{\rho \beta}\ \text{with}\ a\neq\beta,\
      {_\rho\omega}^{(\nu,\beta+1)}_{(\nu,\beta)}=\eta^{1\beta}\eta_{\rho\beta}
      -\eta^{1\nu}\eta_{\rho \nu},\ {_\rho\omega}^{(\nu,\beta+1)}_{(\beta,a)}
      =-\eta^{1a}\eta_{\rho \nu};
    \end{equation}
  \item for $2\leqslant \beta< \nu\leq N$ we set
    \begin{equation}\label{eq:25}
     {_{\lambda\mu}\xi}^{(\nu,\beta+1)}=
    -\eta^{11}(\eta_{\lambda\beta}\eta_{\mu \nu} - \eta_{\lambda
      \nu}\eta_{\mu\beta});
  \end{equation}
  \item for $1<\nu\leqslant N$ and $\beta=1$ we set
    \begin{equation}\label{eq:26}
    {_\lambda\omega}^{(\nu,2)}_{(\nu,\mu)}=1\quad\text{and}\quad
    {_\mu\omega}^{(\nu,2)}_{(\nu,\lambda)}=1.
  \end{equation}
  \end{subequations}
\end{itemize}
\end{theorem}

\begin{proof}
  The particular ordering of equations is important for an inductive reasoning.
  Indeed, we observe that in the case $N=3$ and $N=4$ the statement yields the examples
  that we discussed in the previous Subsections.
  It is the starting point of our proof by induction.
  We assume that the above claim is true in dimension~$N-1$.
  Using~\eqref{eq:51} to calculate the derivatives of~$F$ we have, for $\nu<N$,
  \begin{gather*}
    S_{\lambda\beta\mu\nu}=\eta ^{a b }(F_{a\lambda\beta }F_{b\mu\nu} -
    F_{a\lambda\nu}F_{b\mu\beta})=
    \\
    \sum_{a,b=1}^{N-1}\eta ^{a b }(F_{a\lambda\beta }F_{b\mu\nu} -
    F_{a\lambda\nu}F_{b\mu\beta})+ \sum_{b=1}^{N-1}\eta ^{N b
    }(F_{N\lambda\beta }F_{b\mu\nu} - F_{N\lambda\nu}F_{b\mu\beta})
    \\
    +\sum_{a=1}^{N-1}\eta ^{a N }(F_{a\lambda\beta }F_{N\mu\nu} -
    F_{a\lambda\nu}F_{N\mu\beta})+ \eta ^{NN }(F_{N\lambda\beta }F_{N\mu\nu} -
    F_{N\lambda\nu}F_{N\mu\beta})
    \\=\\
    \sum_{a,b=1}^{N-1}\eta ^{a b }(F_{a\lambda\beta }F_{b\mu\nu} -
    F_{a\lambda\nu}F_{b\mu\beta})+ \sum_{a=1}^{N-1}\eta ^{N a
    }(F_{N\lambda\beta }F_{a\mu\nu} - F_{a\lambda\nu}F_{N\mu\beta})
    \\
    +\sum_{a=1}^{N-1}\eta ^{a N }(F_{a\lambda\beta }F_{N\mu\nu} -
    F_{N\lambda\nu}F_{a\mu\beta})+ \eta ^{NN }(F_{N\lambda\beta }F_{N\mu\nu} -
    F_{N\lambda\nu}F_{N\mu\beta})
    \\=\\
    \sum_{a,b=1}^{N-1}\eta ^{a b }(F_{a\lambda\beta }F_{b\mu\nu} -
    F_{a\lambda\nu}F_{b\mu\beta})+ \eta ^{NN }( u^{(N,\beta)}_\lambda
    u^{(N,\nu)}_\mu - u^{(N,\nu)}_\lambda u^{(N,\beta)}_\mu)
    \\
    +\eta ^{N1}(u^{(N,\beta)}_\lambda\eta_{\mu\nu}-
    \eta_{\lambda\nu}u^{(N,\beta)}_\mu)
    + \sum_{a=2}^{N-1}\eta ^{N a }(u^{(N,\beta)}_\lambda u^{(\nu,a)}_\mu
    - u^{(\nu,a)}_\lambda u^{(N,\beta)}_\mu)
    \\
    +\eta ^{1 N} (\eta_{\lambda\beta }u^{(N,\nu)}_\mu
    -u^{(N,\nu)}_\lambda\eta_{\mu\beta})
    +\sum_{a=2}^{N-1}\eta ^{a N } (u^{(\beta,a)}_\lambda u^{(N,\nu)}_\mu
    - u^{(N,\nu)}_\lambda u^{(\beta,a)}_\mu).
  \end{gather*}
  The first summand can be written in the necessary form by assumption, and it
  is evident the remaining terms do not change the assumption form but
  supplement it by adding $N$th terms to $\psi^\gamma$'s. So, the last
  expression yields part of equations~\eqref{eq:11} and all of~\eqref{eq:18}.

  For $\nu=N$ and $\gamma=(N,\beta+1)$ we have
\begin{gather*}
  S_{\lambda \beta\mu N}=\eta ^{a b } (F_{a \lambda \beta }F_{b\mu N} - F_{a
    \lambda N}F_{b\mu\beta})= \sum_{a =2 }^N\sum_{b=2}^N\eta ^{a b } (F_{a
    \lambda \beta }F_{b\mu N} - F_{a \lambda N}F_{b\mu\beta})
  \\
  +\eta^{11}(F_{1\lambda \beta }F_{1\mu N} - F_{1\lambda N}F_{1\mu\beta})
  +\sum_{b=2 }^{N}\eta ^{1 b } (F_{1\lambda \beta }F_{b\mu N} - F_{1\lambda
    N}F_{b\mu\beta}) +\sum_{a =2}^{N}\eta ^{1 a } (F_{a \lambda \beta }F_{1\mu
    N} - F_{a \lambda N}F_{1\mu\beta})=
  \\
  \sum_{a =2 }^N\sum_{b=2 }^N\eta ^{a b } (u^{(\beta,a)}_\lambda u^{(N,b)}_\mu- u^{(N,b)}_\lambda u^{(\beta,a)}_\mu)
  +\eta^{11}(\eta_{\lambda \beta }\eta_{\mu N} - \eta_{\lambda N}\eta_{\mu\beta}) \\
  +\sum_{a=2}^{N}\eta ^{1 a} \eta_{\lambda \beta}u^{(N,a)}_\mu
  -\sum_{a=2}^{N}\eta ^{1 a} \eta_{\lambda N}u^{(a,\beta)}_\mu
  +\sum_{a=2}^{N}\eta ^{1 a} u^{(a,\beta)}_\lambda\eta_{\mu N}
  -\sum_{a=2}^{N}\eta ^{1 a} u^{(N,a)}_\lambda\eta_{\mu\beta},
\end{gather*}
which yield the remaining coefficients in~\eqref{eq:11} and all of
\eqref{eq:24}, \eqref{eq:25}.

Lastly, the trivial equations $u^{(\lambda,i)}_\mu=u^{(\mu,i)}_\lambda$ give
$\psi^{(i,2)}_{km}=0$ for every value of $k$ and $m$, and
equation~\eqref{eq:26}, completing the proof.
\end{proof}

\begin{corollary}\label{cor:structure}
  Let $N\geq 3$. Then, WDVV equations $S_{\lambda\beta\mu\nu}=0$ can be
  assembled in the form of $(N-1)(N-2)/2$ subsystems of the form
  \eqref{eq:LinearComplexForm}, each of which can be interpreted as a linear
  line congruence in a suitable Pl\"ucker embedding.
\end{corollary}

The Structure Theorem is already an indication that we can extract Hamiltonian
systems from the WDVV equations. However, we need more work to come to that
conclusion.

\subsection{The Metric Theorem}
\label{sec:hamilt-oper-syst}

Here we make a further step in the direction of proving that the linear line
congruences in which the WDVV equations can be arranged
(Corollary~\ref{cor:structure}) give rise, for each choice of an independent
variable~$t^p$, $2\leq p\leq N$, to $N-2$ Hamiltonian systems of $n$
first-order conservation laws, as in the cases $N=3$ and $N=4$, with
third-order homogeneous Hamiltonian operators in canonical form~\eqref{DPHamOp}.

We recall that $n=N(N-1)/2$. In this Subsection, overlined indices run from $1$
to $n$, and have a representation $\bar i=(I,i)$, with $2\leq i\leq I\leq N$.
Additionally, we introduce a `generalized' Kronecker delta: $\delta^P=1$ if the condition $P$ holds, and
$\delta^P=0$ otherwise.

The only missing ingredient in the construction of a Hamiltonian operator $A_2$
as above (see also Subsection~\ref{sec:third-order-operator}) is the scalar
product $\phi$, as in
$A_2^{\bar i\bar j}=\phi_{\bar \alpha\bar \beta}\psi^{\bar \alpha}_{\bar
  i}\psi^{\bar \beta}_{\bar j}$, fulfilling the linear algebraic system 
\begin{subequations}\label{eq:H}
\begin{gather}
\mathcal
H^1_{\bar i\bar j\bar k\bar s}\colon
\phi_{\bar a\bar b}(\psi^{\bar a}_{\bar i\bar s}\psi^{\bar b}_{\bar j\bar k}
+\psi^{\bar a}_{\bar j\bar s}\psi^{\bar b}_{\bar k\bar i}
+\psi^{\bar a}_{\bar k\bar s}\psi^{\bar b}_{\bar i\bar j})=0,
\\
\mathcal
H^2_{\bar i\bar j\bar k}\colon
\phi_{\bar a\bar b}(\omega^{\bar a}_{\bar i}\psi^{\bar b}_{\bar j\bar k}
+\omega^{\bar a}_{\bar j}\psi^{\bar b}_{\bar k\bar i}
+\omega^{\bar a}_{\bar k}\psi^{\bar b}_{\bar i\bar j})=0,
\\
\mathcal
H^3_{\bar i\bar j\bar k}\colon
\phi_{\bar a\bar b}(\theta^{\bar a}_{\bar i}\psi^{\bar b}_{\bar j\bar k}
+\theta^{\bar a}_{\bar j}\psi^{\bar b}_{\bar k\bar i}
+\theta^{\bar a}_{\bar k}\psi^{\bar b}_{\bar i\bar j})=0,
\\
\mathcal H^4_{\bar i\bar k}\colon
\phi_{\bar a\bar b}(\psi^{\bar a}_{\bar i\bar k}\xi^{\bar b}
+\omega^{\bar a}_{\bar k}\theta^{\bar b}_{\bar i}
-\omega^{\bar a}_{\bar i}\theta^{\bar b}_{\bar k})=0,
\end{gather}
\end{subequations}
where $1\leq \bar i< \bar j<\bar k< \bar s\leq n$ and the constants
$\psi^{\bar \gamma}_{\bar k\bar m}$,
$\omega^{\bar \gamma}_{\bar k}:={}_p\omega^{\bar \gamma}_{\bar k}$,
$\theta^{\bar \gamma}_{\bar m}:={}_\mu\omega^{\bar \gamma}_{\bar m}$,
$\xi^{\bar \gamma}:={}_{p\mu}\xi^{\bar \gamma}$ are defined
in the Structure Theorem. We will prove the following Metric Theorem.
\begin{theorem}[Metric Theorem]\label{theor:metric-theorem}
Up to scaling by a constant multiple, there exists a unique solution of the
system~$\mathcal H$,
\[
\phi_{(A,a)(B,b)}=
\eta^{AB}\eta^{(a-1)(b-1)}-\eta^{A(b-1)}\eta^{B(a-1)}.
\]
The matrix $\phi=(\phi_{(A,a)(B,b)})$ coincides, in coordinates, with the
metric induced by $\eta$ on $2$-vectors of $\mathbb{R}^N$, hence it is
invertible.
\end{theorem}

The proof of the above Theorem is a consequence of three lemmas, where we
consecutively solve subsystems of the system~$\mathcal H$,
$\mathcal H^1:=\{\mathcal H^1_{\bar i\bar j\bar k\bar s}\}$, the joint system
of $\mathcal H^2$ and $\mathcal H^3$,
$\mathcal H^2:=\{\mathcal H^2_{\bar i\bar j\bar k}\}$,
$\mathcal H^3:=\{\mathcal H^3_{\bar i\bar j\bar k}\}$, and
$\mathcal H^4:=\{\mathcal H^4_{\bar i\bar k}\}$, which are associated with
different submatrices of the matrix~$\phi$.  Each of these proofs consists of
three parts: induction argument, existence of a solution and its
uniqueness. Induction argument implies that a dimension~$N-1$ system is a
subsystem of a dimension~$N$ system, and therefore the solution set of the
former is a solution subset of the latter. This argument is used to show that
$\phi_{\bar a\bar b}$ as in the statement of Theorem~\ref{theor:metric-theorem}
satisfies a given system. Lastly, we find a subsystem of a given system with a
unique solution.  Together with the existence argument it implies the
uniqueness for the entire system. The result of each lemma is used to prove
subsequent ones.

\begin{remark}
  Several parts of the proofs below require separate consideration of
  lower-dimensional cases, therefore we note here that in dimensions up to~9
  the validity of the claim was verified symbolically. We provide Maple program
  files to support our claims in \cite{opavit26:_maple_reduc_wdvv}.
\end{remark}

\begin{lemma}\label{lemma_H1}
If $\phi$ satisfies the system $\mathcal H^1$,
then up to a scaling by a constant we have
\[
(\phi)_{(A,a)(B,b)}=
\begin{pmatrix}
\eta^{AB}\eta^{(a-1)(b-1)} - \eta^{A(b-1)}\eta^{B(a-1)}
\end{pmatrix}
\]
where $3\leq a\leq A\leq N$, $3\leq b\leq B\leq N$.
\end{lemma}
\begin{proof}
  \textbf{1. Induction reasoning.}  First of all, we note that in dimension
  three the system is vacuous, so below $N>3$.  For $N=3$ and~$N=4$
  Theorem~\ref{theor:metric-theorem} follows the Examples discussed above in
  this Section.  Also, since $\psi^{(N,m)}_{(I,i)(J,j)}=0$ for $I,J<N$, the
  equation~$\mathcal H^1_{\bar i\bar j\bar k\bar s}=0$ in dimension $N-1$
  coincides with its counterpart in dimension~$N$,
  \begin{gather*}
\sum\limits_{2\leq a\leq A\leq N}\sum\limits_{2\leq b\leq B\leq N} \phi_{(A,a)(B,b)}(\psi^{(A,a)}_{\bar i\bar s}\psi^{(B,b)}_{\bar j\bar k}
+\psi^{(A,a)}_{\bar j\bar s}\psi^{(B,b)}_{\bar k\bar i}
+\psi^{(A,a)}_{\bar k\bar s}\psi^{(B,b)}_{\bar i\bar j})
=\\
\sum\limits_{2\leq a\leq A\leq N-1}\sum\limits_{2\leq b\leq B\leq N-1} \phi_{(A,a)(B,b)}(\psi^{(A,a)}_{\bar i\bar s}\psi^{(B,b)}_{\bar j\bar k}
+\psi^{(A,a)}_{\bar j\bar s}\psi^{(B,b)}_{\bar k\bar i}
+\psi^{(A,a)}_{\bar k\bar s}\psi^{(B,b)}_{\bar i\bar j}),
\end{gather*}
and thus the solution set of
  the dimension~$N-1$ system is a solution subset of its counterpart in
  dimension~$N$, which allows us to use a dimensional induction.

  \textbf{2. Existence.} In general, expanding
  $\mathcal H^1_{\bar i\bar j\bar k\bar s}$ reduces to expanding sums of the
  form $\sum\limits_{a,A}\phi_{(A,a)\bar b}\psi^{(A,a)}_{(I,i)(J,j)}$, which is
  equal to
\begin{gather}
\begin{split}\label{eq:PsiExpansion}
\sum\limits_{A=2}^N\sum_{a=2}^A\phi_{(A,a)\bar c}\psi^{(A,a)}_{(I,i)(J,j)}=
\phi_{(i,j+1)\bar b}\eta^{IJ}\delta^{J>I}\delta^{i>j}
+\phi_{(I,j+1)\bar b}\eta^{iJ}\delta^{I>i}\delta^{I>j}
-\phi_{(J,i+1)\bar b}\eta^{jI}
\\
-\phi_{(j,i+1)\bar b}\eta^{IJ}\delta^{J>j}\delta^{j>i}
-\phi_{(j,I+1)\bar b}\eta^{iJ}\delta^{I>i}\delta^{J>j}\delta^{j>I}
-\phi_{(J,I+1)\bar b}\eta^{ij}\delta^{J>I}\delta^{I>i}.
\end{split}
\end{gather}
Let us give a brief explanation of this formula.  From the Structure Theorem we
know that
$\psi^{(\nu,\beta+1)}_{(\nu,a)(\beta,b)}=-\psi^{(\nu,\beta+1)}_{(\beta,b)(\nu,a)}
=\eta^{ab}$ are the only nonzero elements of the matrix~$\psi^{(\nu,\beta+1)}$.
Therefore, we can pick out all nonzero elements of the above sum manually.
For example, if $A=I$ then $a$ must be equal to either~$J$ or~$j$. But if $J=j$
then there is only one nonzero term, and hence the Kronecker deltas. The form of
the equation~$\mathcal H^1_{\bar i\bar j\bar k\bar s}$ is completely determined
by the mutual configuration%
\footnote{Recall that in general capital parameters are greater or equal than
  their small counterparts and $I\leq J\leq K\leq S$, but the configuration of
  small parameters~$(i,j,k,s)$, and small and capital parameters with different
  symbols, e.g. $(j,I)$, are unknown.}  of the elements of
tuple~$(i,j,k,s,I,J,K,S)$, that is, if we change a value of a parameter without
changing the mutual configuration the corresponding equations and their
solutions can be obtained one from another by a simple substitution.

For induction reasons we assume that found~$\phi$ for all parameters with $S<N$
satisfy the system.  Given an equation in dimension~$N-1$ and increasing all
values in the associated tuple of parameters by one, and thus not changing the
configuration of parameters, we obtain a dimension~$N$ equation whose solution
satisfies Lemma by construction. Unfortunately, such construction does not
cover dimension~$N$ equations simultaneously containing~$2$ and~$N$ in their
parameters (as the function~$\rk$ is not defined at~1 for any of its
arguments). Therefore, one should modify the construction to avoid the
necessity to raise the value of two, and it is done by making the induction
base big enough.

In the most general configuration, there are eight different parameters in the
tuple.  Let us take $N=9$ to be the base case. For $N>9$, if both~two and~$N$
are in the tuple, according to the Dirichlet's (pigeonhole) principle, we can
change enough values of parameters to preserve the configuration without
changing the value of~two. Indeed, with at least nine holes (in dimension ten)
and only eight pigeons, there is a hole without pigeons. Denoting this hole
by~$\alpha$, we can obtain any dimension ten equation from a dimension nine
equation by increasing values of all parameters greater than or equal
to~$\alpha$, while preserving the values less than~$\alpha$.

\textbf{3. Uniqueness.}  Let us choose three distinct numbers
$2\leqslant \alpha<\beta<\gamma\leqslant N$ and form six numbers
$(\gamma,\alpha)$, $(\gamma,\beta)$, $(\gamma,\gamma)$, $(\beta,\alpha)$,
$(\beta,\beta)$, $(\alpha,\alpha)$.
We can select a subsystem of 15~equations by picking out four numbers,
which are ordered in an increasing order to become indices of the equation $\mathcal H^1$,
out of the above six.  For example,
\begin{gather*}
  \mathcal H^1_{(\alpha,\alpha)(\beta,\alpha)(\beta,\beta)(\gamma,\alpha)}=
  \phi_{(A,a)(B,b)}\psi^{(A,a)}_{(\gamma,\alpha)(\alpha,\alpha)}\psi^{(B,b)}_{(\beta,\beta)(\beta,\alpha)}
  \\
  -\phi_{(A,a)(B,b)}\psi^{(A,a)}_{(\gamma,\alpha)(\beta,\alpha)}\psi^{(B,b)}_{(\beta,\beta)(\alpha,\alpha)}
  +\phi_{(A,a)(B,b)}\psi^{(A,a)}_{(\gamma,\alpha)(\beta,\beta)}\psi^{(B,b)}_{(\beta,\alpha)(\alpha,\alpha)}=
  \\
  \phi_{(\gamma,\alpha+1)(\beta,\alpha+1)}\eta^{\alpha\alpha}\eta^{\beta\beta}
  +\phi_{(\beta,\alpha+1)(\beta,\alpha+1)}\eta^{\alpha\gamma}\eta^{\beta\alpha}
  -\phi_{(\gamma,\beta+1)(\beta,\alpha+1)}\eta^{\alpha\alpha}\eta^{\beta\alpha}
  \\
  -\phi_{(\gamma,\alpha+1)(\beta,\alpha+1)}\eta^{\alpha\beta}\eta^{\beta\alpha}
  -\phi_{(\beta,\alpha+1)(\beta,\alpha+1)}\eta^{\beta\gamma}\eta^{\alpha\alpha}
  +\phi_{(\gamma,\beta+1)(\beta,\alpha+1)}\eta^{\beta\alpha}\eta^{\alpha\alpha}
  =\\
  \phi_{(\gamma,\alpha+1)(\beta,\alpha+1)}(\eta^{\alpha\alpha}\eta^{\beta\beta}-\eta^{\alpha\beta}\eta^{\beta\alpha})
  +\phi_{(\beta,\alpha+1)(\beta,\alpha+1)}(\eta^{\alpha\gamma}\eta^{\beta\alpha}-\eta^{\beta\gamma}\eta^{\alpha\alpha})=
  \\
  \phi_{(\gamma,\alpha+1)(\beta,\alpha+1)}M_{(\beta,\alpha+1)(\beta,\alpha+1)}
  -\phi_{(\beta,\alpha+1)(\beta,\alpha+1)}M_{(\gamma,\alpha+1)(\beta,\alpha+1)}.
\end{gather*}
  The equations of this subsystem have the form
\[
\phi_{l_1l_2}M_{l_3l_4}-\phi_{l_3l_4}M_{l_1l_2}=0,
\]
where
$l_1,l_2,l_3,l_4\in\{(\gamma,\alpha+1),(\gamma,\beta+1),(\beta,\alpha+1)\}$
with $l_1\leqslant l_2$, $l_3\leqslant l_4$ and for any appropriate indices
$M_{(A,a+1)(B,b+1)}:=\eta^{AB}\eta^{ab}-\eta^{Ab}\eta^{Ba}$.  It can be solved
to
\begin{gather*}
\begin{pmatrix}
\phi_{(\beta,\alpha+1)(\beta,\alpha+1)} & \phi_{(\gamma,\alpha+1)(\beta,\alpha+1)} & \phi_{(\gamma,\alpha+1)(\gamma,\alpha+1)} &
\phi_{(\gamma,\beta+1)(\beta,\alpha+1)} & \phi_{(\gamma,\beta+1)(\gamma,\alpha+1)} & \phi_{(\gamma,\beta+1)(\gamma,\beta+1)} \\
M_{(\beta,\alpha+1)(\beta,\alpha+1)} & M_{(\gamma,\alpha+1)(\beta,\alpha+1)} & M_{(\gamma,\alpha+1)(\gamma,\alpha+1)} &
M_{(\gamma,\beta+1)(\beta,\alpha+1)} & M_{(\gamma,\beta+1)(\gamma,\alpha+1)} & M_{(\gamma,\beta+1)(\gamma,\beta+1)}
\end{pmatrix}
\\=
\begin{pmatrix}
xa_1 &xa_2 &xa_3 &xa_4 &xa_5 &xa_6 \\
ya_1 &ya_2 &ya_3 &ya_4 &ya_5 &ya_6
\end{pmatrix}
\end{gather*}
where $a_i\neq0$ for $i=1,\dots,6$. Since $\beta<N$, we can use the induction
argument,
$\phi_{(\beta,\alpha+1)(\beta,\alpha+1)}=M_{(\beta,\alpha+1)(\beta,\alpha+1)}$
up to a constant, to show that $x=y$.

The found $\phi$'s do not exhaust all unknowns, and therefore we now pick four
distinct numbers $2\leqslant \alpha<\beta<\gamma<\delta\leqslant N$ and form
10~different numbers as above. Then we select the following three associated
equations:
\begin{gather*}
\mathcal H^1_{(\alpha,\alpha)(\beta,\alpha)(\gamma,\alpha)(\delta,\alpha)}=
-\phi_{(\delta,\alpha+1)(\beta,\alpha+1)}\eta^{\alpha\alpha}\eta^{\gamma\alpha}
+\phi_{(\delta,\alpha+1)(\gamma,\alpha+1)}\eta^{\alpha\alpha}\eta^{\beta\alpha}
+\phi_{(\delta,\alpha+1)(\gamma,\beta+1)}\eta^{\alpha\alpha}\eta^{\alpha\alpha}
\\
+\phi_{(\delta,\beta+1)(\gamma,\alpha+1)}\eta^{\alpha\alpha}\eta^{\alpha\alpha}
-\phi_{(\delta,\beta+1)(\gamma,\alpha+1)}\eta^{\alpha\alpha}\eta^{\alpha\alpha}
-\phi_{(\delta,\alpha+1)(\gamma,\alpha+1)}\eta^{\alpha\beta}\eta^{\alpha\alpha}
\\
-\phi_{(\delta,\gamma+1)(\beta,\alpha+1)}\eta^{\alpha\alpha}\eta^{\alpha\alpha}
+\phi_{(\delta,\alpha+1)(\beta,\alpha+1)}\eta^{\alpha\alpha}\eta^{\gamma\alpha}
+\phi_{(\delta,\gamma+1)(\beta,\alpha+1)}\eta^{\alpha\alpha}\eta^{\alpha\alpha}
=\\
\left(\phi_{(\delta,\alpha+1)(\gamma,\beta+1)}
+\phi_{(\delta,\beta+1)(\gamma,\alpha+1)}
-\phi_{(\delta,\beta+1)(\gamma,\alpha+1)}\right)(\eta^{\alpha\alpha})^2=0;
\\[1ex]
\mathcal H^1_{(\alpha,\alpha)(\beta,\beta)(\gamma,\gamma)(\delta,\delta)}=
\phi_{(\delta,\alpha+1)(\gamma,\beta+1)}\eta^{\delta\alpha}\eta^{\gamma\beta}
-\phi_{(\delta,\beta+1)(\gamma,\alpha+1)}\eta^{\delta\beta}\eta^{\gamma\alpha}
+\phi_{(\delta,\gamma+1)(\beta,\alpha+1)}\eta^{\delta\gamma}\eta^{\beta\alpha}=0;
\\[1ex]
\mathcal H^1_{(\alpha,\alpha)(\beta,\beta)(\gamma,\alpha)(\delta,\delta)}=
\phi_{(\delta,\alpha+1)(\gamma,\beta+1)}\eta^{\delta\alpha}\eta^{\alpha\beta}
-\phi_{(\delta,\beta+1)(\gamma,\alpha+1)}\eta^{\delta\beta}\eta^{\alpha\alpha}
+\phi_{(\delta,\gamma+1)(\beta,\alpha+1)}\eta^{\delta\alpha}\eta^{\beta\alpha}
\\
+\phi_{(\delta,\alpha+1)(\beta,\alpha+1)}(\eta^{\delta\gamma}\eta^{\beta\alpha}-\eta^{\delta\alpha}\eta^{\gamma\beta})
=\\
+\phi_{(\delta,\alpha+1)(\gamma,\beta+1)}\eta^{\delta\alpha}\eta^{\alpha\beta}
-\phi_{(\delta,\beta+1)(\gamma,\alpha+1)}\eta^{\delta\beta}\eta^{\alpha\alpha}
+\phi_{(\delta,\gamma+1)(\beta,\alpha+1)}\eta^{\delta\alpha}\eta^{\beta\alpha}
\\
+(\eta^{\delta\beta}\eta^{\alpha\alpha}-\eta^{\delta\alpha}\eta^{\beta\alpha})(\eta^{\delta\gamma}\eta^{\beta\alpha}-\eta^{\delta\alpha}\eta^{\gamma\beta})=0
\end{gather*}
and observe that all equations in the subsystem obtained by the above choice of
indices are, up to a multiplier, coincident with one in the above triple.  We
can solve these three equations (and thus the entire system), ignoring
nondegeneracy issues, with respect to
$\phi_{(\delta,\alpha+1)(\gamma,\beta+1)}$,
$\phi_{(\delta,\beta+1)(\gamma,\alpha+1)}$ and
$\phi_{(\delta,\gamma+1)(\beta,\alpha+1)}$ and the solution is exactly as in
the statement of Lemma. On the other hand, for construction reasons, all
$\eta$'s in this system belong to the same submatrix~$M_{\alpha,\beta,\gamma,\delta;\alpha,\beta,\gamma,\delta}$
formed by retaining $\alpha$th, $\beta$th, $\gamma$th and $\delta$th rows and columns of the matrix~$\eta$
and by rejecting the other rows and columns,
which could be identically zero in a large dimension without triggering the degeneracy of~$\eta$
(in this case, the values~$\phi$'s associated with minors must thus vanish as well,
which does not contradict the statement). Therefore, we need more equations for these unknowns.

Let $\mathcal E = \mathbb Z\cap[2,N]\setminus\{\alpha,\beta,\gamma,\delta\}$.
Assuming $N\geq9$ (and in a lesser dimension symbolic computations show validity of Metric Theorem),
there exist~$\rho_1,\rho_2,\rho_3\in\mathcal E$ such that $\eta^{\rho_1\delta}\eta^{\rho_2\gamma}\eta^{\rho_3\beta}\neq0$
since out of four columns of~$\eta$ only one can almost vanish
(we do not take into consideration the first element of a column).
Without loss of generality we assume $\eta^{\rho\delta}=0$ for all $\rho\in\mathcal E$
(and more generally for all $\rho>1$ taking into account the previous assumptions).
Let us consider the system
\begin{gather*}
\mathcal H^1_{(\rho_2,\alpha)(\beta,\rho_1)(\gamma,\gamma)(\delta,\delta)}\colon
\eta^{\delta\rho_2}\eta^{\gamma\rho_1}\phi_{(\delta,\alpha+1)(\gamma,\beta+1)}-\eta^{\delta\rho_1}\eta^{\gamma\rho_2}\phi_{(\delta,\beta+1)(\gamma,\alpha+1)}=0;
\\[1ex]
\mathcal H^1_{(\rho_3,\alpha)(\beta,\beta)(\gamma,\rho_1)(\delta,\delta)}\colon
\eta^{\delta\rho_3}\eta^{\beta\rho_1}\phi_{(\delta,\alpha+1)(\gamma,\beta+1)}+\eta^{\delta\rho_1}\eta^{\beta\rho_3}\phi_{(\delta,\gamma+1)(\beta,\alpha+1)}=0;
\\[1ex]
\mathcal H^1_{(\rho_3,\alpha)(\beta,\beta)(\gamma,\gamma)(\delta,\rho_2)}\colon
\eta^{\rho_3\beta}\eta^{\gamma\rho_2}\phi_{(\delta,\gamma+1)(\beta,\alpha+1)}-\eta^{\rho_2\beta}\eta^{\gamma\rho_3}\phi_{(\delta,\beta+1)(\gamma,\alpha+1)}=0.
\end{gather*}
If $\phi_{(\delta,\alpha+1)(\gamma,\beta+1)}=0$,
then also $\phi_{(\delta,\gamma+1)(\beta,\alpha+1)}=0$ and $\phi_{(\delta,\beta+1)(\gamma,\alpha+1)}=0$
and we have achieved our goal.
Otherwise, interchanging $\rho_1$ and $\rho_2$ in $\mathcal
H^1_{(\rho_2,\alpha)(\beta,\rho_1)(\gamma,\gamma)(\delta,\delta)}$
we have
\begin{displaymath}
  \mathcal
  H^1_{(\rho_1,\alpha)(\beta,\rho_2)(\gamma,\gamma)(\delta,\delta)}\colon
  \eta^{\delta\rho_1}\eta^{\gamma\rho_2}\phi_{(\delta,\alpha+1)(\gamma,\beta+1)}-\eta^{\delta\rho_2}\eta^{\gamma\rho_1}\phi_{(\delta,\beta+1)(\gamma,\alpha+1)}=0,
\end{displaymath}
where the first summand is nonzero, and thus all the factors of the second one
as well.  Similarly, from the equation
$\mathcal H^1_{(\rho_1,\alpha)(\beta,\beta)(\gamma,\rho_3)(\delta,\delta)}$ we
deduce $\eta^{\delta\rho_3}\eta^{\beta\rho_1}\neq0$.  Finally, the system
\begin{gather*}
  \mathcal H^1_{(\rho_1,\alpha)(\beta,\rho_1)(\gamma,\gamma)(\delta,\delta)}\colon
  \eta^{\delta\rho_1}\eta^{\gamma\rho_1}(\phi_{(\delta,\alpha+1)(\gamma,\beta+1)}-\phi_{(\delta,\beta+1)(\gamma,\alpha+1)})=0;
  \\[1ex]
  \mathcal H^1_{(\rho_1,\alpha)(\beta,\beta)(\gamma,\rho_1)(\delta,\delta)}\colon
  \eta^{\delta\rho_1}\eta^{\beta\rho_1}(\phi_{(\delta,\alpha+1)(\gamma,\beta+1)}+\phi_{(\delta,\gamma+1)(\beta,\alpha+1)})=0;
  \\[1ex]
  \mathcal H^1_{(\rho_1,\alpha)(\beta,\beta)(\gamma,\gamma)(\delta,\rho_1)}\colon
  \eta^{\rho_1\beta}\eta^{\gamma\rho_1}(\phi_{(\delta,\gamma+1)(\beta,\alpha+1)}-\phi_{(\delta,\beta+1)(\gamma,\alpha+1)})=0
\end{gather*}
evidently admits only zero solutions.

It remains to show that the found two sets of $\phi$'s,
\begin{gather*}
\{\phi_{(\beta,\alpha+1)(\beta,\alpha+1)}, \phi_{(\gamma,\alpha+1)(\beta,\alpha+1)}, \phi_{(\gamma,\alpha+1)(\gamma,\alpha+1)},
\phi_{(\gamma,\beta+1)(\beta,\alpha+1)}, \phi_{(\gamma,\beta+1)(\gamma,\alpha+1)}, \phi_{(\gamma,\beta+1)(\gamma,\beta+1)}\}
\\ \text{and}\
\{\phi_{(\delta,\alpha+1)(\gamma,\beta+1)},\phi_{(\delta,\beta+1)(\gamma,\alpha+1)},\phi_{(\delta,\gamma+1)(\beta,\alpha+1)}\},
\end{gather*}
exhaust all unknowns.  Let us use the induction argument again and concentrate
on dimension~$N$ assuming we have a result in dimension~$N-1$.  Then the first
set of $\phi$'s (with $\gamma=N$) gives all diagonal elements of the
matrix~$\phi$, $\phi_{(N,x)(N,x)}$, $x=3,\dots,N$, as well as
$\phi_{(N,\alpha+1)(\beta,\alpha+1)}$, $\phi_{(N,\beta+1)(\beta,\alpha+1)}$,
$\phi_{(N,\beta+1)(N,\alpha+1)}$.  In the $(N,x)$th row for a fixed $x<N$ (we
take $\beta<N-1$ here) this set gives one $\phi_{(N,x)(\beta,x)}$, with
$\beta\geqslant x$, and $\phi_{(N,x)(x-1,\alpha)}$, with $\alpha\leqslant x-1$.
What it omits is elements $\phi_{(N,x)(\beta,y)}$ with $y>x$ and
$\phi_{(N,x)(y,\alpha)}$ with $y<x$.  But these are exactly the elements, which
are respectively given by $\phi_{(\delta,\alpha+1)(\gamma,\beta+1)}$ and
$\phi_{(\delta,\gamma+1)(\beta,\alpha+1)}$ of the second set ($\delta=N$).  In
the $(N,N)$th row the first set gives (now $\beta=N-1$)
$\phi_{(N,N)(N-1,\alpha+1)}$ and $\phi_{(N,N)(N,\alpha+1)}$, and thus omits
$\phi_{(N,N)(y,\alpha+1)}$, $y<N-1$, but it is given by
$\phi_{(\delta,\gamma+1)(\beta,\alpha+1)}$ for $\gamma<N-1$.  The value
$\gamma=N-1$ covers the omitted above case $\beta=N-1$.
\end{proof}

Having solved the system~$\mathcal H^1$ we can proceed to the next subsystem of~$\mathcal H$.
\begin{lemma}\label{H2_H3_solns}
Up to a scaling by a constant multiple, there is a unique solution to the joint system of equations $\mathcal H^2$ and $\mathcal H^3$,
\[
\phi_{(A,a)(B,2)}=\eta^{AB}\eta^{(a-1)1} - \eta^{A1}\eta^{B(a-1)}
\]
where $3\leqslant a\leqslant A\leqslant N$, $2\leqslant B\leqslant N$.
\end{lemma}

\begin{proof}
\textbf{1. Induction reasoning.}
Since $\psi^{\bar a}_{(B,2)\bar c}$=0 for all appropriate~$\bar a,B,\bar c$,
this system contains only unknowns of the form~$\phi_{(A,2)(B,b)}$ where $b>2$
(given that the system~$\mathcal H^1$ is already solved).
Applying Structure Theorem to an equation~$\mathcal H^2_{\bar i\bar j\bar k}$ we notice
that the expanded expression
\begin{equation}
\begin{split}\label{eq:OmegaThetaExpansion}
\sum_{A=3}^N\sum_{a=3}^A\phi_{(A,a)\bar b}\ \omega^{(A,a)}_{(K,k)}&=
\sum_{x=2}^{k-1}\phi_{(k,x+1)\bar b}\eta^{1K}\eta_{p x}\delta^{K>k}
+\sum_{x=2}^{K-1}\phi_{(K,x+1)\bar b}\eta^{1k}\eta_{p x}\\
&
-\sum_{x=K+1}^{N}\phi_{(x,K+1)\bar b}\eta^{1k}\eta_{p x}
-\sum_{x=k+1}^{N}\phi_{(x,k+1)\bar b}\eta^{1K}\eta_{p x}\delta^{K>k}
\end{split}
\end{equation}
depends on lower-indices $\eta$, which are the functions of~$\eta^{N*}$.
(Recall that $\psi^{\bar a}_{(B,2)\bar c}$=0 for all appropriate~$\bar a,B,\bar c$,
and therefore we consider sums starting with~three).
On the other hand, the entire expression contains only incomplete sums as
\begin{gather*}
\left(\sum_{x=2}^{k-1}\eta^{xj}\eta_{p x}
+\sum_{x=k+1}^{N}\eta^{xj}\eta_{p x}\right)\eta^{IJ}\eta^{ki}\delta^{J>I}\delta^{i>j}\eta^{1K}\delta^{K>k},
\end{gather*}
which we can reduce to
\begin{gather*}
\eta^{ki}\eta^{IJ}\delta^{J>I}\delta^{i>j}\eta^{1K}\delta^{K>k}
(\delta^{j=p}-\eta^{1j}\eta_{p 1}-\eta^{kj}\eta_{p k}).
\end{gather*}
Collecting further coefficients of~$\eta_{1p}$, $\eta_{ip}$, $\eta_{jp}$
and~$\eta_{kp}$ we notice that they all vanish, which leaves only
$\eta_{**}$-independent terms.  The detailed lengthy computation is given in an
auxiliary file \cite{opavit26:_maple_reduc_wdvv}.  The net result is the form
of the equations~$\mathcal H^2_{\bar i\bar j\bar k\bar s}$
and~$\mathcal H^3_{\bar i\bar j\bar k\bar s}$, for which a separate
consideration is not necessary since elements of $\omega$ and $\theta$ are the
same up to a value of parameter, are completely determined by the mutual
configuration of the elements of tuple~$(i,j,k,I,J,K)$.

\textbf{2. Existence.} In dimensions greater than~7 we can use the same argument as in the existence part of Lemma~\ref{lemma_H1}
to prove that the desired solution satisfies the systems~$\mathcal H^2$ and~$\mathcal H^3$.

\textbf{3. Uniqueness.}  To show the uniqueness of the solution we find a
subsystem with a unique solution.  Let us choose two distinct numbers
$2\leqslant \alpha<\beta\leqslant N$ and three associated numbers
$(\alpha,\alpha)$, $(\beta,\alpha)$, $(\beta,\beta)$. Then if
$\mu\notin\{\alpha,\beta\}$ we have
\begin{gather*}
\mathcal H^3_{(m,\mu)(\alpha,\alpha)(\beta,\alpha)}=
(\phi_{(m,2)(\beta,\alpha+1)}-\eta^{\beta m}\eta^{1\alpha}+\eta^{\beta \alpha}\eta^{1m})\eta^{\alpha\alpha},\\
\mathcal H^3_{(m,\mu)(\alpha,\alpha)(\beta,\beta)}=
(\phi_{(m,2)(\beta,\alpha+1)}-\eta^{\beta m}\eta^{1\alpha}+\eta^{\beta \alpha}\eta^{1m})\eta^{\alpha\beta},\\
\mathcal H^3_{(m,\mu)(\beta,\alpha)(\beta,\beta)}=
(\phi_{(m,2)(\beta,\alpha+1)}-\eta^{\beta m}\eta^{1\alpha}+\eta^{\beta \alpha}\eta^{1m})\eta^{\beta\beta}.
\end{gather*}
Here $2\leqslant m\leqslant N$ and $m$ is taken to be less than~$\alpha$, the
expressions for other~$m$ are similar.  Thus, if at least one of
$\eta^{\alpha\alpha}$, $\eta^{\beta\alpha}$, $\eta^{\beta\beta}$ does not
vanish we have
$\phi_{(m,2)(\beta,\alpha+1)}=\eta^{\beta m}\eta^{1\alpha}-\eta^{\beta
  \alpha}\eta^{1m}$.  Otherwise, the equations
$\mathcal H^3_{(m,\mu)(\alpha,\alpha)(\beta,\gamma)}$ for
$\gamma\in\mathbb Z\cap[2,N]\setminus\{\alpha,\beta\}$ reduce to
$\eta^{\alpha\gamma}\phi_{(m,2)(\beta,\alpha+1)}=0$, and the equations
$\mathcal H^3_{(m,\mu)(\beta,\gamma)(\beta,\beta)}$ for
$\gamma\in\mathbb Z\cap[2,N]\setminus\{\alpha,\beta\}$ reduce to
$\eta^{\beta\gamma}\phi_{(m,2)(\beta,\alpha+1)}=0$.  At least one of $\eta$'s
involved is nonzero in view of nondegeneracy of~$\eta$, and thus
$\phi_{(m,2)(\beta,\alpha+1)}=\eta^{\beta m}\eta^{1\alpha}-\eta^{\beta
  \alpha}\eta^{1m}=0$ once again.

Analogously for $p\notin\{\alpha,\beta\}$ the equations
$\mathcal H^2_{(m,p)(\alpha,\alpha)(\beta,\alpha)}$,
$\mathcal H^2_{(m,p)(\alpha,\alpha)(\beta,\beta)}$,
$\mathcal H^2_{(m,p)(\beta,\alpha)(\beta,\beta)}$ give rise to the same three
equation as above. Thus, we found all $\phi_{(m,2)(\beta,\alpha+1)}$ for
$2\leq\alpha<\beta$, leaving out only~$\phi_{(m,2)3}$ since
$(\beta,\alpha)>3$ in all the cases above. To find them, one has to solve a
single equation $\mathcal H^2_{(\alpha,p)(\beta,\mu)(m,2)}$ for each~$m$ and
$2<\alpha,\beta<N$ such that $\eta^{\alpha\beta}\neq0$, which obviously exists
due to nondegeneracy of~$\eta$ (notice $\alpha=m$ gives a vacuous equation).
\end{proof}

For the system~$\mathcal H^4$ we present the direct proof (giving crucial
details only; complete computation is provided in the auxiliary files
\cite{opavit26:_maple_reduc_wdvv}) for a simple reason.  After proving the
induction step, which is more complicated than before since it requires using
the Jacobi and the Laplace identities to get rid of $(\eta_{**})$-dependence,
we end up with a few remaining equations, which can be dealt directly.

\begin{lemma}
Up to a constant multiple, the system~$\mathcal H^4$ has a unique solution
\[
\phi_{(A,2)(B,2)}=\eta^{AB}\eta^{11}-\eta^{A1}\eta^{B1},
\]
where $2\leqslant A,B\leqslant N$.
\end{lemma}

\begin{proof}
  We shall recall some terminology and elementary results from the theory of
  determinants; they will be used throughout the proof. Given a square
  matrix~$M$, its rejector minor~$M_{i_1i_2\dots i_r,j_1j_2\dots j_r}$ is the
  determinant of the matrix formed by deleting $i_1$th, \dots $i_r$th rows and
  $j_1$,\dots $j_r$th columns of the matrix~$M$; its retainer
  minor~$N_{i_1i_2\dots i_r,j_1j_2\dots j_r}$ is the determinant of the matrix
  formed by retaining $i_1$th, \dots $i_r$th rows and $j_1$th,\dots $j_r$th
  columns of the matrix~$M$ and deleting the rest; the rejector
  cofactor~$A_{i_1i_2\dots i_r,j_1j_2\dots j_r}$ is the signed rejector
  minor~$M_{i_1i_2\dots i_r,j_1j_2\dots j_r}$,
  $A_{i_1i_2\dots i_r,j_1j_2\dots j_r}=(-1)^kM_{i_1i_2\dots i_r,j_1j_2\dots
    j_r}$ where $k=\sum_{s=1}^r(i_s+j_s)$;
  see~\cite[pp. 18--20]{vein99:_deter_their_applic_mathem_physic}.

  We expand all parameters in
  $\mathcal H^4_{\bar i\bar k}\colon\phi_{\bar a\bar b} (\psi^{\bar a}_{\bar
    i\bar k}\xi^{\bar b}+\omega^{\bar a}_{\bar k}\theta^{\bar b}_{\bar
    i}-\omega^{\bar a}_{\bar i}\theta^{\bar b}_{\bar k})=0$ using the Structure
  Theorem. First, using~\eqref{eq:OmegaThetaExpansion} and the previous lemmas,
  similarly to Lemma~\ref{H2_H3_solns} we reduce
  $\phi_{\bar a\bar b}(\omega^{\bar a}_{\bar k}\theta^{\bar b}_{\bar
    i}-\omega^{\bar a}_{\bar i}\theta^{\bar b}_{\bar k})$ to cyclic sums of the
  form
\begin{gather*}
  \sum_{b\in\mathbb Z\cap[2,N]\setminus\{k,K\}}\sum_{a\in\mathbb
    Z\cap[2,N]\setminus\{I\}}\eta^{Ik}\eta^{ab}\eta^{1i}\eta_{pa}\eta^{K1}\eta_{\mu b}\delta^{K>k}
\end{gather*}
where we use the identity $\eta^{xy}\eta_{yz}=\delta^{x=z}$ up to two times
(as explained in the proof of the previous lemma)
to get rid of the incomplete sums in favour of its missing terms as a result of a
straightforward but lengthy computation. The remaining term of~$\mathcal H^4_{\bar i\bar j}$,
$\phi_{\bar a\bar b}\psi^{\bar a}_{\bar i\bar k}\xi^{\bar b}$, upon using the
previous lemmas and the Structure Theorem reduces to
\begin{gather*}
-\sum_{b=3}^N\sum_{a=2}^{b-1}N_{Ik,ba}\eta^{iK}
\eta^{11}R_{p\mu,ba}\delta^{K>I}\delta^{I>k}
-\sum_{b=3}^N\sum_{a=2}^{b-1}N_{ik,ba}\eta^{IK}
\eta^{11}R_{p\mu,ba}\delta^{K>I}\delta^{I>i}\delta^{i>k}
\\
+\sum_{b=3}^N\sum_{a=2}^{b-1}N_{ki,ba}\eta^{KI}
\eta^{11}R_{p\mu,ba}\left(\delta^{K>I}\delta^{I>i}\delta^{K>k}\delta^{k>i}
+\delta^{I=K}\delta^{k>i}\delta^{K>k}\right)
\\
+\sum_{b=3}^N\sum_{a=2}^{b-1}N_{KI,ba}
\eta^{ki}\eta^{11}R_{p\mu,ba}\delta^{K>I}
-\sum_{b=3}^N\sum_{a=2}^{b-1}N_{Ik,ba}\eta^{iK}
\eta^{11}R_{p\mu,ba}\delta^{I=K}\\
+\sum_{b=3}^N\sum_{a=2}^{b-1}N_{Ki,ba}\eta^{kI}
\eta^{11}R_{p\mu,ba}\eta_{\mu a}\left(\delta^{K>I}\delta^{I>i}
+\delta^{I=K}\right)
+\sum_{b=3}^N\sum_{a=2}^{b-1}N_{kI,ba}\eta^{Ki}
\eta^{11}R_{p\mu,ba}\delta^{K>I}\delta^{K>k}\delta^{k>I},
\end{gather*}
where $N_{ab,cd}$ is the retainer minor of the matrix~$(\eta^{ij})$,
and $R_{p\mu,ba}$ is the retainer minor of the matrix~$(\eta_{ij})$.
Using the Jacobi identity for complementing minors,
\[
R_{p\mu,ba}=\eta_{p a}\eta_{\mu b}-\eta_{p b}\eta_{\mu a} = \frac1{\det \eta} A_{p\mu,ab},
\]
where $A_{p\mu,ab}$ is the rejector cofactor of~$(\eta^{ij})$, after which,
complementing the sum to the indices $1\leqslant a<b\leqslant N$ we use the
Laplace sum formula
\[
\sum_{c<d} N_{ab,cd}A_{p\mu,cd}=\delta^{a=p}\delta^{b=\mu}\det \eta.
\]
In this way, we reduce the whole $\psi\xi$-term to
\begin{gather*}
\delta^{K=\mu}\delta^{i=p}\eta^{11}\eta^{kI}
+\sum_{x=1}^NN_{Ki,1x}\eta^{11}R_{p\mu,1x}\eta^{kI}
-\sum_{x=1}^NN_{ik,1x}\eta^{11}R_{p\mu,1x}\eta^{KI}\delta^{K>I}\delta^{i>k}
\\
+\delta^{I=\mu}\delta^{k=p}\eta^{11}\eta^{Ki}\delta^{K>I}\delta^{I>i}(\delta^{K>k}\delta^{k>I}-\delta^{I>k})
+\sum_{x=1}^NN_{Ik,1x}\eta^{11}R_{p\mu,1x}\eta^{iK}(\delta^{K=I}\delta^{K>k}-\delta^{K>I}\delta^{I>i}\delta^{I>k})
\\
+\delta^{k=\mu}\delta^{i=p}\eta^{11}\eta^{KI}\delta^{K>k}\delta^{k>i}
+\sum_{x=1}^NN_{ki,1x}\eta^{11}R_{p\mu,1x}\eta^{KI}\delta^{K>k}\delta^{k>i}
-\eta^{11}\eta^{KI}\delta^{K>I}\delta^{i>k}\delta^{i=\mu}\delta^{k=p}
\\
-\sum_{x=1}^NN_{kI,1x}\eta^{11}R_{p\mu,1x}\eta^{Ki}\delta^{K>I}\delta^{I>i}\delta^{K>k}\delta^{k>I}
-\sum_{x=1}^NN_{KI,1x}\eta^{11}R_{p\mu,1x}\eta^{ki}\delta^{K>I}\delta^{I>i}.
\end{gather*}
Here we can again use the summation formula above to kill the sums,
after which combining the remaining terms with $\omega\theta$-terms above
we deduce that the final nonvanishing equations are
\begin{gather*}
\mathcal H^4_{(I,p)(K,\mu)}=-\phi_{(I,2)(K,2)}+\eta^{11}\eta^{KI}-\eta^{1K}\eta^{1I},\\
\mathcal H^4_{(I,\mu)(K,p)}=\phi_{(I,2)(K,2)}+\eta^{1I}\eta^{K1}-\eta^{11}\eta^{1K},\\
\mathcal H^4_{(\mu,p)(K,p)}=\phi_{(K,2)(2,2)}-\eta^{11}\eta^{2K}+\eta^{12}\eta^{1K},\\
\mathcal H^4_{(p,p)(\mu,p)}=-\phi_{(2,2)(2,2)}+\eta^{11}\eta^{22}-(\eta^{12})^2,
\end{gather*}
which gives the desired result.
\end{proof}

For a better understanding of the proofs below we present the following matrix
\[
\begin{pmatrix}
\phi^4_{11} & \phi^4_{12} & \phi^{2,3}_{13} & \phi^4_{14} & \phi^{2,3}_{15} & \phi^{2,3}_{16} & \phi^4_{17} & \phi^{2,3}_{18} & \phi^{2,3}_{19} & \phi^{2,3}_{1\,10} \\[1ex]
\phi^4_{21} & \phi^4_{22} & \phi^{2,3}_{23} & \phi^4_{24} & \phi^{2,3}_{25} & \phi^{2,3}_{26} & \phi^4_{27} & \phi^{2,3}_{28} & \phi^{2,3}_{29} & \phi^{2,3}_{2\,10} \\[1ex]
\phi^{2,3}_{31} & \phi^{2,3}_{32} & \phi^1_{33} & \phi^{2,3}_{34} & \phi^1_{35} & \phi^1_{36} & \phi^{2,3}_{37} & \phi^1_{38} & \phi^1_{39} & \phi^1_{3\,10} \\[1ex]
\phi^4_{41} & \phi^4_{42} & \phi^{2,3}_{43} & \phi^4_{44} & \phi^{2,3}_{45} & \phi^{2,3}_{46} & \phi^4_{47} & \phi^{2,3}_{48} & \phi^{2,3}_{49} & \phi^{2,3}_{4\,10} \\[1ex]
\phi^{2,3}_{51} & \phi^{2,3}_{52} & \phi^1_{53} & \phi^{2,3}_{54} & \phi^1_{55} & \phi^1_{56} & \phi^{2,3}_{57} & \phi^1_{58} & \phi^1_{59} & \phi^1_{5\,10} \\[1ex]
\phi^{2,3}_{61} & \phi^{2,3}_{62} & \phi^1_{63} & \phi^{2,3}_{64} & \phi^1_{65} & \phi^1_{66} & \phi^{2,3}_{67} & \phi^1_{68} & \phi^1_{69} & \phi^1_{6\,10} \\[1ex]
\phi^4_{71} & \phi^4_{72} & \phi^{2,3}_{73} & \phi^4_{74} & \phi^{2,3}_{75} & \phi^{2,3}_{76} & \phi^4_{77} & \phi^{2,3}_{78} & \phi^{2,3}_{79} & \phi^{2,3}_{7\,10} \\[1ex]
\phi^{2,3}_{81} & \phi^{2,3}_{82} & \phi^1_{83} & \phi^{2,3}_{84} & \phi^1_{85} & \phi^1_{86} & \phi^{2,3}_{87} & \phi^1_{88} & \phi^1_{89} & \phi^1_{8\,10} \\[1ex]
\phi^{2,3}_{91} & \phi^{2,3}_{92} & \phi^1_{93} & \phi^{2,3}_{94} & \phi^1_{95} & \phi^1_{96} & \phi^{2,3}_{97} & \phi^1_{98} & \phi^1_{99} & \phi^1_{9\,10} \\[1ex]
\phi^{2,3}_{10\, 1} & \phi^{2,3}_{10\, 2} & \phi^1_{10\, 3} & \phi^{2,3}_{10\, 4} & \phi^1_{10\, 5} & \phi^1_{10\, 6} & \phi^{2,3}_{10\, 7} & \phi^1_{10\, 8} & \phi^1_{10\, 9} & \phi^1_{10\, 10}
\end{pmatrix}
\]
It shows (in dimension~5) that the entry $\phi^i_{\bar j\bar k}$ is a part of
solution set of the system~$\mathcal H^i$ (which includes the joint
system~$\mathcal H^2$ and~$\mathcal H^3$). To put it into another perspective,
we introduce a sequence $a_m:=(m,2)$, $m\geq2$,
$a=\{1,2,4,7,11,16,22,29,\dots\}$, which is the first row of the
matrix~$(f_{\lambda im})$ in Subsection~\ref{sec:wdvv-systems}. If both the row
and the column of the entry~$\phi_{ab}$ belongs to this sequence (\emph{i.e.}
$\phi_{\bar a\bar b}=\phi_{(A,2)(B,2)}$) it comes as a solution of the
system~$\mathcal H^4$, if only one (\emph{i.e.}
$\phi_{\bar c\bar d}=\phi_{(A,2)(B,b)}$) - of the joint system~$\mathcal H^2$
and~$\mathcal H^3$ , if none (\emph{i.e.}
$\phi_{\bar e\bar f}=\phi_{(A,a)(B,b)}$) - of the system~$\mathcal H^1$.

\subsection{The Hamiltonian Theorem}

Let us choose two indices $\lambda$, $\mu$ such that $2\leq \lambda<\mu\leq N$.
Then, according to the Structure Theorem, the subsystem of WDVV equations
$S_{\lambda\beta\mu\nu}=0$ can be written as
\begin{equation}
  \label{eq:7}
  {}_\lambda\psi^\alpha_{I} u^I_\mu = {}_\lambda Z^\alpha_{\mu},
\end{equation}
where
${}_\lambda\psi^\alpha_{I}={}_\lambda\psi^\alpha_{IJ}u^J_\lambda + {}_\lambda
\omega^\alpha_I$ and
${}_\lambda Z^\alpha_{\mu}= {_\mu\omega}^\alpha_I u^I_\lambda +
{}_{\lambda\mu}\xi^\alpha$.  We also know that there exists a scalar
product~$\phi$ that fulfills the equations~\eqref{eq:H}. We need to make a
final step before the Hamiltonian property can be proved: nondegeneracy of
$\psi=({}_\lambda\psi^\alpha_{I})$. We will use the following simple Proposition.
\begin{proposition}
  Let $\phi$ be a nondegenerate symmetric bilinear form on a $k$-dimensional
  real vector space $A$. Let $\{A^\alpha\mid \alpha=1,\ldots,n\}$ be generators
  of $A$. If the Gram matrix $(\phi(A^\alpha,A^\beta))$ is nondegenerate, then the generators $\{A^\alpha\}$ are independent.
\end{proposition}
\begin{proof}
  The map $\cdot^\flat\colon A\to A^*$, defined by $\phi$, is a linear
  isomorphism. If, say, $A^1 = \sum_{\beta=2}^n k_\beta A^\beta$ then
  \begin{equation*}
    (A^1)^\flat = \sum_{\beta=2}^n k_\beta (A^\beta)^\flat.
  \end{equation*}
  Computing the above covector relation on the vectors in
  $\{A^\alpha\mid \alpha=1,\ldots,n\}$, \emph{e.g.}
  $(A^1)^\flat(A^\beta)=\phi(A^1,A^\beta)$ and similarly on the right-hand
  side, it turns out that the first row of the Gram matrix is a linear
  combination of the other rows.
\end{proof}

\begin{corollary}\label{theor:nondegen_psi}
  The matrix $\psi=({}_\lambda\psi^\alpha_{I})$ is nondegenerate on a dense
  open subspace of the space of dependent variables $(u^I_\lambda )$.
\end{corollary}
\begin{proof}
  It is a direct consequence of the Metric Theorem, which states that $\phi$
  coincides, in coordinates, with the (nondegenerate) metric induced naturally
  by $\eta$ on $\wedge^2\mathbb{R}^N$. Note that $A^\alpha_\lambda$
  (see~\eqref{eq:38}) can be written as
  $A^\alpha_\lambda = {}_\lambda\psi^\alpha_{I} \mathrm du^I_\lambda$.
\end{proof}
Note that the fact that $\psi$ can be degenerate on a singular subset is due to
the identification of $u^a$ with a basis element $e^a$ in view of their
transformation under linear changes of variables
(Subsection~\ref{sec:hamilt-struct-wdvv}). The identification does not hold
where the coordinates become zero.

\begin{theorem}[Hamiltonian theorem]\label{theor:HamTheorem}
  Let $N\geq 3$, and choose an independent variable $t^p$, ${2\leq p\leq
  N}$. Then, provided we exclude a singular set, we obtain $N-2$ Hamiltonian
  systems of first-order conservation laws, indexed by $\mu$ such that
  $2\leq \mu\leq N$ and $\mu\neq p$:
  \begin{equation}\label{eq:57}
    (u^I_p)_\mu = ({_pV}^I_\mu)_p,\qquad I=1,\ldots,n,
  \end{equation}
  where the fluxes are functions of $(u^J_p)$:
  ${_pV}^I_\mu = {_pV}^I_\mu(u^J_p)$.  Each of the above systems is Hamiltonian
  with respect to one and the same third-order Hamiltonian operator $A_2$ in
  canonical form \eqref{DPHamOp} defined by the Monge metric~\ref{eq:30}
  ${_pf}_{ij}=\phi_{\alpha\beta}\, {_p\psi}^\alpha_i{_p\psi}^\beta_j$, where ${_p\psi}^\alpha_i$
  is defined as above and
  $\phi_{\alpha\beta}$ is defined in the Metric Theorem~\ref{theor:metric-theorem}.
\end{theorem}
\begin{proof}
  First of all, we observe that the structure theorem requires $p < \mu$
  and $1\leq \beta < \nu$ in $S_{p \beta \mu \nu}$.  However, if we take
  $p > \nu$, then we should consider the equations
  $S_{\mu \nu p \beta}$ with $\nu<\beta$ in view of the symmetry
  in~\eqref{eq:81}.

  The fluxes of the systems are defined by:
  \begin{equation}\label{eq:10}
    {_pV}^I_\mu = {_p\psi}^I_\alpha\, {_pZ}^\alpha_\mu,
  \end{equation}
  where $({_p\psi}^I_\alpha) = ({_p\psi}^\alpha_I)^{-1}$ on a dense open subset of the
  space of dependent variables. The conditions for the above system to be
  Hamiltonian with respect to~$A_2$ are, respectively, the third and fourth
  equation of system~\eqref{eq:H}, and are fulfilled in view of the Metric
  Theorem.
\end{proof}

\section{Reducing WDVV equations}

In this section we will study the problem of the determination of nontrivial
WDVV equations, and then if it is possible to write the reduced WDVV system in
passive orthonomic form. See also the Subsection~\ref{sec:first-probl-determ}
of the Introduction. The solution of this problem has a direct impact on the
relationship between the Hamiltonian systems in the Hamiltonian
Theorem~\ref{theor:HamTheorem}: indeed, if WDVV can be written in passive
orthonomic form, then the systems are commuting.

\begin{theorem}[Reduction theorem]\label{th:reduc-wdvv-equat}
  Let $N\geq 3$, and consider the WDVV equations with the quasihomogeneity
  condition in the \emph{semisimple} case, in which all homogeneity weights
  are distinct. Choose an independent variable $t^p$, $2\leq p\leq
  N$.   Then:
  \begin{enumerate}
  \item the following subsystem of the WDVV equation~\eqref{eq:5}
  \begin{equation}\label{eq:35}
    S_L = \{S_{p\beta\mu\nu} = 0\mid 1\leq\beta<\nu\leq N,\
    2\leq \mu\leq N,\ \mu\neq p\}
  \end{equation}
  is equivalent, on an open subset of the space of dependent variables, to a
  system of $\binom{N}{3}$ partial differential equations of the form
  \begin{equation}
    \label{eq:2}
    f_{ijk} = G_{ijk}(f_{plm}),\quad i,j,k\in\{2,\ldots,N\}\setminus\{p\},
  \end{equation}
  where $G_{ijk}(f_{plm})$ are rational functions whose denominator is
  $\det({_p\psi}^\alpha_i)$;
\item all other equations in the WDVV equation~\eqref{eq:5} vanish on account
  of equations~\eqref{eq:2}.
\end{enumerate}
\end{theorem}
\begin{proof}
  By the Structure Theorem, for any fixed $\mu\neq p$ the $\mu$-subsystem
  \begin{displaymath}
    \{S_{p\beta\mu\nu} = 0\mid 1\leq\beta<\nu\leq N\}
  \end{displaymath}
  of the system~\eqref{eq:35} can be put in the form
  ${_p\psi}^\alpha_I V^I_\mu = {_pZ}^\alpha_\mu$. That implies that the $\mu$-subsystem can
  locally (where $({_p\psi}^\alpha_I)$ is invertible) be solved with respect to the
  variables $(u^I_p)$, and the solution is ${_pV}^I_\mu = {_p\psi}^I_\alpha \,{_pZ}^\alpha_\mu$.

  Note that for each third-order derivative of the form $f_{ijk}$, where
  $i$, $j$, $k\in\{2,\ldots,N\}\setminus\{p\}$, there is at least one index
  $\mu$ such that ${_pV}^I_\mu = f_{ijk}$ for a certain value of~$I$. So, all
  third-order derivatives which do not contain~$p$ as an index are expressed as
  rational functions with $\det({_p\psi}_I^\alpha)$ in the denominator.

  Two problems arise immediately.
  \begin{enumerate}
  \item It can happen, for some pairs of indices~$J$ and~$K$, that ${_pV}^J_\mu$
    and ${_pV}^K_{\mu'} $ correspond to the same third-order derivative, which can
    be an element of $(u^I_p)$ or not. Are the expressions of the solutions the
    same?
  \item Provided the previous problem is solved in the affirmative, do all other
    equations in the full system $S_{\alpha\beta\gamma\nu}=0$ identically
    vanish on account of the solutions of~\ref{eq:35}?
  \end{enumerate}

  Let us recall~\cite{dubrovin06:_encyc_mathem_physic} that the space of all
  $N$-dimensional semisimple Frobenius manifolds is
  $N(N-1)/2$-dimensional. That implies that, at each fixed point
  $(t^\lambda,f,f_{\tau_1},f_{\tau_2})\in J_2E$ the fiber of the WDVV equation
  $\mathcal{E}\subset J_3E$ should contain a space which is parametrized by at
  least $N(N-1)/2$ free parameters.

  Now, let us suppose that for two indices $J$ and $K$, the fluxes ${_pV}^J_\mu$
  and ${_pV}^K_{\mu'} $ correspond to the same third-order derivative. If that
  derivative is one of the elements of the vector $(u^I_p)$, then the two
  fluxes define the same trivial equation in the sense of
  Proposition~\ref{pro:trivialeq}. If that derivative is not in $(u^I_p)$ we
  have two rational expressions of the coordinates $(u^I_p)$ for
  the same third-order derivative. They are the same: if they would be
  different, they would introduce further constraints on the fibre of the WDVV
  equation over a fixed second-order jet, so to lower the dimension which is
  prescribed above.

  Finally, all other equations in the WDVV equation~\eqref{eq:5} must vanish on
  account of~\eqref{eq:2} for the same dimensional reason as above.
\end{proof}
\begin{remark}
  The above statement has been verified by symbolic computations when $N=4$ and
  $\eta$ is arbitrary and when $N=5$ and $\eta$ is one of Dubrovin's canonical
  forms, see \cite{opavit26:_maple_reduc_wdvv}.
\end{remark}

It is then a simple consequence to observe that each equation in~\eqref{eq:2}
is solved with respect to a principal derivative which is distinct from others,
and no principal derivatives appear in any of the right-hand sides of the
equations.
\begin{corollary}
  Let $N\geq3$. Under the hypotheses of semisimplicity, WDVV equations for
  quasihomogeneous potentials $F$ can be written locally in orthonomic form.
\end{corollary}

In full generality, provided we reduced the WDVV equations to~\eqref{eq:2}, we
are able to connect the compatibility of the WDVV equations
to the commutativity of the $N-2$ first-order WDVV systems. Here, compatibility
has to be meant in the sense that cross-derivatives of the equations do not
turn into new integrability conditions.

\begin{proposition}\label{theor:Commutativity} Let $N\geq 3$ and $\eta$ be
  arbitrary. Suppose that WDVV equations can be (locally) reduced to the
  orthonomic system~\eqref{eq:2}.  Then, the following conditions are
  equivalent:
  \begin{itemize}
  \item WDVV equations are compatible;
  \item the $N-2$ first-order
  systems defined in the Hamiltonian Theorem~\ref{theor:HamTheorem} (see
  also~\eqref{eq:16}) are commuting;
  \item WDVV equations in orthonomic form are passive.
  \end{itemize}
\end{proposition}

\begin{proof}
  Let us fix $p$ such that $2\leq p\leq N$. Let $\mu$ and $\kappa$ be such that
  $2\leq \mu, \kappa\leq N$ and different from~$p$. Denote
  $v^m:=u^m_\mu$, $w^m:=u^m_\kappa$; these are functions of $u^m:=u^m_p$. For
  ease of notation, let us denote ${}_p\psi^\gamma_i =
  \psi^\gamma_{ki}u^k + {}_p\omega^\gamma_i$.
  Thus, the WDVV equations inter alia give rise to three subsystems
\begin{gather}
(\psi^\gamma_{km}u^k+{_p\omega}^\gamma_m)v^m-{_\mu\omega}^\gamma_mu^m-{_{p\mu}\xi}^\gamma=0\label{eq:CommutingFlows1},\\
(\psi^\gamma_{km}u^k+{_p\omega}^\gamma_m)w^m-{_\kappa\omega}^\gamma_mu^m-{_{p\kappa}\xi}^\gamma=0\label{eq:CommutingFlows2},\\
(\psi^\gamma_{km}w^k+{_\kappa\omega}^\gamma_m)v^m-{_\mu\omega}^\gamma_mw^m-{_{\kappa\mu}\xi}^\gamma=0\label{eq:CommutingFlows3}.
\end{gather}
Differentiating~\eqref{eq:CommutingFlows1} with respect to~$t^\kappa$ and using
the first-order systems $(u^m)_\mu = (v^m)_p$, $(u^m)_\mu = (w^m)_p$,  we have
\begin{gather*}
\psi^\gamma_{km}u^k_{\kappa}v^m+(\psi^\gamma_{km}u^k+{_p\omega}^\gamma_m)v^m_{\kappa}-{_\mu\omega}^\gamma_mu^m_{\kappa}=
\\
\psi^\gamma_{km}w^k_{p}v^m+(\psi^\gamma_{km}u^k+{_p\omega}^\gamma_m)v^m_{\kappa}-{_\mu\omega}^\gamma_mw^m_{p}=0.
\end{gather*}
Analogously, differentiating~\eqref{eq:CommutingFlows2} with respect to~$t^\mu$ gives
\[
\psi^\gamma_{km}v^k_{p}w^m+(\psi^\gamma_{km}u^k+{_p\omega}^\gamma_m)w^m_{\mu}-{_\kappa\omega}^\gamma_mv^m_{p}=0,
\]
with their difference being
\[
\psi^\gamma_{km}w^k_{p}v^m+(\psi^\gamma_{km}u^k+{_p\omega}^\gamma_m)v^m_{\kappa}-{_\mu\omega}^\gamma_mw^m_{p}
-\psi^\gamma_{km}v^k_{p}w^m-(\psi^\gamma_{km}u^k+{_p\omega}^\gamma_m)w^m_{\mu}+{_\kappa\omega}^\gamma_mv^m_{p}=0,
\]
which in view of the differential consequence of~\eqref{eq:CommutingFlows3},
\[
\psi^\gamma_{km}w^k_{p}v^m-\psi^\gamma_{km}v^k_{p}w^m+{_\kappa\omega}^\gamma_mv^m_{p}-{_\mu\omega}^\gamma_mw^m_p=0,
\]
simplifies to
\[
(\psi^\gamma_{km}u^k+{_p\omega}^\gamma_m)(v^m_{\kappa}-w^m_{\mu})=0.
\]
Taking advantage of the nondegeneracy of~$({}_p\psi^\gamma_i)$ (again, on
a dense open subset of the space of dependent variables) we get the condition
$v^m_{\kappa}-w^m_{\mu}=0$. The vanishing of this condition, for every
$3\leq \mu, \kappa\leq N$, is equivalent to the compatibility, in view of the
above calculations; on the other hand, it is clearly equivalent to the
passivity of the orthonomic form~\eqref{eq:2}.

Computing the derivatives we have
\begin{equation}
  \label{eq:32}
  v^m_{\kappa} = u^i_\kappa\pd{v^m}{u^i} = w^i_pv^m_i = w^i_ju^j_{t^p}v^m_i,
\end{equation}
where we denoted by $u^j_{t^p}$ the first-order jet space coordinate. Hence,
the condition $v^m_{\kappa}-w^m_{\mu}=0$ is equivalent to the commutativity of
the two first-order systems that we considered above.
\end{proof}

\begin{corollary}\label{cor:wdvv-equat-sys}
  Without loss of generality, let us choose $p=2$, and denote $u^I=u^I_2$,
  $I=1$, \dots, $n$. Suppose that one of the equivalent conditions of the
  previous Proposition holds.  Then, the restriction of the
  equations~\eqref{eq:8} to the prolonged WDVV equations in the above
  orthonomic form:
  \begin{equation}
    \label{eq:13}
    \mathcal{E}^{(1)}\subset J_4E \subset \hat{J}_4E \subset J_1J_3E
  \end{equation}
  is identified, on an open dense subset of the space of dependent variables,
  by $N-2$ Hamiltonian first-order systems of conservation laws of the form
  \begin{equation}
    \label{eq:16}
    (u^I)_3 = (u^I_3)_x,\  \ldots,\  (u^I)_N = (u^I_{N})_x,\qquad I=1,\ldots,n
  \end{equation}
  where all fluxes are functions of the same variables: $u^I_3=u^I_3(u^j)$,
  \dots, $u^I_N=u^I_N(u^j)$.  The further conditions
  $(u^I_\lambda)_\mu = (u^I_\mu)_\lambda$, $2<\lambda<\mu\leq N$, vanish
  identically.
\end{corollary}
\begin{proof}
  The systems in~\eqref{eq:16} are all Hamiltonian, according with the
  Hamiltonian Theorem~\ref{theor:HamTheorem}. Repeating the same reasoning as
  in Subsection~\ref{sec:main-example-sys}, Proposition~\ref{pro:passive} and
  Proposition~\ref{pro:commuting}, we get the result.
\end{proof}

Unfortunately, we have no result of the type of Proposition~\ref{pro:passive} in
arbitrary dimension $N$. In other words, we cannot state that the local
orthonomic form of WDVV is also passive. But, thanks to computer algebra, we
can state the following Theorem.

\begin{theorem}\label{prop:reduc-wdvv}
  Let $N=4$ and $N=5$, and let $\eta$ be one of the two canonical forms
  $\eta^{(1)}$ or $\eta^{(2)}$ (see the Introduction). Then, the orthonomic
  form of the WDVV system is also passive; equivalently, the first-order
  systems of conservation laws commute.
\end{theorem}

Our computational results make us conjecture that the statement of the above
Theorem has a general validity.

\begin{conjecture}[Passivity Conjecture]
  In the semisimple case, the WDVV system can be written locally in passive
  orthonomic form.
\end{conjecture}

We end this Section by a straightforward, but important, statement, concerning
the equivalence between solutions of WDVV equations and solutions of the $N-2$
systems in the case when they are commuting.

\begin{proposition}\label{pro:reduc-wdvv-equat}
  Let us choose an independent variable $t^p$, and assume that the $N-2$
  systems of Corollary~\ref{cor:wdvv-equat-sys} are commuting. Then, from any
  joint solution $(u^I)$ of the two systems~\eqref{eq:42} it is possible to
  obtain a solution of the WDVV equations up to second degree polynomials.
\end{proposition}
\begin{proof}
  The proof is a direct generalization of the proof of
  Proposition~\ref{pro:equiv-wdvv-4}.
\end{proof}

\section{WDVV equations as bi-Hamiltonian systems}
\label{sec:wdvv4_HamOp}

It is natural to ask ourselves if the third-order Hamiltonian structure that we
found for all WDVV equations (in the form of first-order systems of
conservation laws) can be complemented with a compatible Hamiltonian operator,
so to form a bi-Hamiltonian pair.

Indeed, as we already mentioned in the Introduction, the simplest WDVV
equation~\eqref{eq:WDVV:simple} can be written in the form of a bi-Hamiltonian
system of first-order conservation laws. The corresponding bi-Hamiltonian pair
is of the form that we discussed in the Introduction \cite{FGMN97}. More
bi-Hamiltonian pairs in dimension $N=3$ were found in
\cite{kalayci98:_alter_hamil_wdvv,kalayci97:_bi_hamil_wdvv}.

More generally, in \cite{vasicek21:_wdvv_hamil} it was proved that, when $N=3$,
the WDVV equation in the form of a system of first-order conservation laws
always admits a bi-Hamiltonian pair of the above mentioned type. We have shown
that systems with similar bi-Hamiltonian pairs are an interesting object of
study \cite{OpanasenkoVitolo2024}.

In $N=4$ the first bi-Hamiltonian pair was discovered in~\cite{PV15} for
$\eta=\eta^{(1)}$. No other bi-Hamiltonian pairs were discovered until now.

In~\cite{lorenzoni25:_compat_hamil} the following Theorem is proved.

\begin{theorem}[\cite{lorenzoni25:_compat_hamil}]\label{theor:compat13}
  Let $A_1$ be a first-order Hamiltonian operator of Ferapontov type (as
  in~\eqref{FerHamOpGen}) that is compatible with a third-order homogeneous
  Hamiltonian operator~$A_2$ in canonical form~\eqref{DPHamOp}. Then, the
  leading coefficient of~$A_1$ fulfills
\begin{equation}\label{eq:232}
    g^{ij} = \psi^i_\gamma Z^{\gamma j} + \psi^j_\gamma Z^{\gamma i}
    -c^{\alpha\beta}w^i_\alpha w^j_\beta,
\end{equation}
where
\begin{itemize}
\item $Z^{\gamma j} = \eta^{\gamma j}_k u^k + \xi^{\gamma j}$ define the family
$V^{i,j}=\psi^i_\gamma Z^{\gamma j}$, indexed by $j$, of fluxes of systems of
conservation laws that are Hamiltonian with respect to $A_2$;
\item the coefficients $w^i_{\alpha j}u^j_x$ of the nonlocal part of $A_1$ have
  the form $w^i_{\alpha j}u^j_x = (w^i_\alpha)_x$ and are a family of fluxes of
  systems of conservation laws that are Hamiltonian with respect to $A_2$.
\end{itemize}
\end{theorem}

We already proved that, having chosen an independent variable $t^p$, all WDVV
first-order systems of conservation laws are Hamiltonian with respect to a
third-order Hamiltonian operator (Hamiltonian Theorem~\ref{theor:HamTheorem}).

Unfortunately, we do not have a similar result concerning a first-order
compatible operator. But, as we have many examples of bi-Hamiltonian WDVV
systems (see, for example, \cite{vasicek21:_wdvv_hamil}) for particular values
of $\eta$, it is interesting to prove that such examples propagate to the orbit
of the action of the invariance group of the WDVV equations.

More precisely, it is known~\cite{dubrovin06:_encyc_mathem_physic} that
the WDVV system admits as a group of symmetries the group of linear
transformations that preserve the unit vector field~$\p_{t^1}$:
\begin{equation*}
  \tilde{t}^\alpha = P^\alpha_\beta t^\beta + Q^\alpha,\qquad
  \det(P^\alpha_\beta)\neq 0,\quad P^\alpha_1 = \delta^\alpha_1,\quad
  \alpha,\beta=1,\dots,N.
\end{equation*}
This group obviously changes the value of the matrix~$\eta$.

\begin{proposition}\label{prop:invariance} A symmetry transformation of the WDVV
  equations preserves the form of the bi-Hamiltonian pair in a bi-Hamiltonian
  first-order WDVV system.
\end{proposition}
\begin{proof}
  The matrix $P=(P^\alpha_\beta)$ of the change of coordinates can be
  factorized as
  \begin{gather*}
    P= T_1\cdot T_2,\qquad \text{where}
    \\
    P=
    \begin{pmatrix}
      1 & (P^1_j)
      \\
      0 & (P^i_j)
    \end{pmatrix},\quad T_1=
    \begin{pmatrix}
      1 & 0
      \\
      0 & (P^i_j)
    \end{pmatrix},\quad T_2=
    \begin{pmatrix}
      1 & P^1_j
      \\
      0 & I
    \end{pmatrix},
  \end{gather*}
  where~$(P^1_i)$ is a row of length~$N-1$, $P^i_j$ is an invertible square
  matrix of order~$N-1$ and~$I$ is the identity matrix.

  The transformation~$T_2$ does not change third-order derivatives of~$f$,
  so we can restrict our attention to~$T_1$.

  It is already known~\cite{vasicek21:_wdvv_hamil} that invariance
  transformations that involve only two independent variables preserve the form
  of the Hamiltonian operators and that~$T_1$ maps first-order WDVV systems
  into new first-order systems.

  The proof is thus completed if we can decompose any invertible matrix into a
  sequence of $2\times 2$~blocks.  And indeed, any matrix in
  $\mathrm{GL}(\mathbb{C}^{N-1})$ can be generated by means of Gauss'
  elementary matrices (up to permutations).
\end{proof}

\begin{remark}
  We tacitly assumed that the pair of Hamiltonian operators remains compatible
  after the transformation $T_1$. Indeed, the Schouten bracket has a geometric
  definition, and we do expect that compatibility is preserved by
  transformations.  Computational experiments suggest that this is the case,
  and the recent results in
  \cite{lorenzoni23:_miura_poiss,lorenzoni25:_compat_hamil} (see also the above
  Theorem~\ref{theor:compat13}) seem to confirm it.
\end{remark}

\begin{remark}
  The transformation $T_2$ does not bring any first-order WDVV system into
  another first-order WDVV system in a direct way. Indeed, our way to obtain
  families of commuting WDVV systems is attached to a given set of independent
  coordinates and to the choice of one of them. We will analyze the
  transformation properties of first-order WDVV systems under the symmetry
  group of WDVV equations in the near future.
\end{remark}

Since in all examples of first-order WDVV systems that we have been able to
deal with (by means of the computational resources and algorithms at our
disposal) we found a bi-Hamiltonian pair $A_1$, $A_2$ of WDVV-type, we can
formulate a Conjecture.

In the general case we have $N-2$ commuting flows \emph{provided} we
  can put the WDVV equations in passive orthonomic form: they are
given by the fluxes of the systems of the Hamiltonian
Theorem~\ref{theor:HamTheorem}. Moreover, the flow $\delta^i_ju^j_x\pd{}{u^i}$
is trivially commuting with all others. So, the dimension of the matrix that
defines the nonlocal part of $A_1$ must be taken as $N-1$. Let us define
$\tilde{\eta}=(\eta^{ij})_{i,j\in\{2,\ldots,N\}}$.

\begin{conjecture}[Bi-Hamiltonian conjecture]
  Every first-order WDVV system admits a unique first-order Hamiltonian
  operator $A_1$ of Ferapontov type~\eqref{FerHamOpGen} that is compatible with
  the third-order Hamiltonian operator from the Hamiltonian
  Theorem~\ref{theor:HamTheorem}.

  The constants $c^{\alpha\beta}$ are defined by
  $(c^{\alpha\beta})_{\alpha,\beta=1}^{N-1}=\tilde\eta \det \tilde\eta$.  In
  particular, $A_1$ is a local operator if and only if $\tilde\eta$ is
  degenerate.
\end{conjecture}

We stress that we formulated the conjecture on the basis of several
computational experiments in dimensions up to $N=5$. More details can be found
in \cite{lorenzoni25:_compat_hamil}.

\bigskip

\textbf{Acknowledgements.} We thank S. Agafonov, P. Antonini, G. Borot,
R. Chiriv\`\i, S. Galkin, E.V.~Ferapontov, B.~Kruglikov, P.~Lorenzoni and
S. Shadrin for useful discussions.  We also thank E. Fasanelli and F. Ricciardi
(INFN, Section of Lecce) for their assistance with the server that we used for
the calculations.


\providecommand{\cprime}{\/{\mathsurround=0pt$'$}}
  \providecommand*{\SortNoop}[1]{}

\end{document}